\documentclass[11pt,twocolumn]{article}

\usepackage{graphicx} 
\usepackage{caption}
\usepackage{newtxtext}
\usepackage{newtxmath}

\usepackage{amsmath}
\usepackage{hyperref}
\usepackage{cite}
\usepackage[version=4]{mhchem}
\usepackage[a4paper,top=1.4cm,bottom=1.9cm,left=1.7cm,right=1.7cm]{geometry}

\usepackage{siunitx}
\usepackage{comment}

\title{\textbf{A Review of Wearable Sweat Monitoring Platforms:\\
From Biomarker Detection to Signal Processing Systems}}

\author{\textbf{Yuhan Zheng}\\[0.5em]
Texas A\&M University\\[0.3em]
\texttt{yuhanz@tamu.edu}
}
\date{}

\begin{document}

\maketitle

\begin{abstract}
Wearable electronics hold great potential in defining new paradigms of modern healthcare, including personalized health management, precision medicine, and athletic performance optimization. This stems from their ability in enabling continuous, real-time health monitoring. To enable molecular-level analysis, biofluids rich in molecular analytes have become one of the most important target samples for wearable sensors. Among them, sweat stands out as an ideal candidate for next-generation wearable health monitoring platforms due to its completely noninvasive nature and ease of acquisition. In recent years, several studies have demonstrated feasible prototype designs for sweat-based wearable sensors. However, one of the major gaps toward large-scale commercialization is the development of clinically validated standards for sweat analysis. One key requirement is to establish the relationship between sweat analytes and those of blood, the latter serving as the gold standard in modern diagnostics. This review provides an overview of sweat biomarkers, with a particular focus on their partitioning mechanisms, which reveal the underlying connections between sweat analytes and their counterparts within systemic metabolic pathways. In addition, this review offers a mechanistic-level examination of biosensors employed in sweat sensing, addressing a gap that has not been adequately covered in prior reviews. Given the critical role of electronic systems in constructing highly integrated wearable sweat-monitoring platforms, this review also analyzes the electronic architectures used for sensor signal processing from an interdisciplinary perspective, with particular emphasis on the analog circuitry that interfaces with electrochemical sensors.
\end{abstract}


\section{Introduction}
Wearable devices can continuously or intermittently acquire information from the body surface and communicate with external systems to enable real-time monitoring or feedback. Among the various sensing modalities, biofluids serve as an important target for achieving molecular-level detection and analysis. Existing studies have shown that the biofluids widely used for wearable biosensors include interstitial fluid (ISF), saliva, tears, and sweat \cite{Heikenfeld2019,Min2021,Ray2019}. These biofluids are preferred because they are common, relatively accessible, and rich in various biomarkers that can reflect the physiological state of the human body. However, these biofluids differ in terms of analyte composition and content, as well as ease of acquisition and collection, which leads to corresponding wearable health monitoring platforms being in different stages of application and having different development prospects.

ISF-based wearable health monitoring platforms currently have the broadest range of applications, which is mainly attributed to their abundance of biomarkers and the relative ease of acquisition through minimally invasive methods. The analytes in ISF come primarily from blood, which traverses continuous capillaries from blood to ISF through three pathways: transcellular diffusion, paracellular transport, and transcytosis. The synergistic action of multiple transport pathways enables both small and large analytes to enter the ISF. Meanwhile, the high surface area-to-volume ratio, high density, and slow flow rate of the capillaries feeding the ISF, together with the driving force generated by blood pressure, facilitate the free exchange of analytes between blood and ISF. As a result, the concentrations of many analytes in ISF are close to those of blood, which is one of the key advantages of ISF \cite{Heikenfeld2019}. Regarding ISF acquisition, two primary approaches are currently employed: reverse iontophoresis (RI) and epidermal microneedle devices \cite{Min2021}. RI applies a weak DC current between two electrodes placed on the skin surface. Under the influence of the electric field, charged analytes (e.g., electrolyte ions) migrate toward the skin surface, while the resulting electro-osmotic flow \cite{Pikal2001} simultaneously drives the transport of neutral analytes (e.g., glucose), thereby allowing non-invasive access to analytes in ISF. However, although RI is non-invasive, it introduces issues such as skin irritation, low analyte concentrations, and strong dependence on skin conditions, which has limited its practical adoption in recent years \cite{Zheng2023}. Instead, microneedle devices implanted under the skin are the mainstream approach for accessing ISF at present. The most representative application of ISF-based wearable sensing platforms is the continuous glucose monitoring system, which has undergone more than two decades of development since its first introduction in 1999. This technology is now relatively mature and has become an important tool for blood glucose management in diabetic patients \cite{Hirsch2018}. From the perspective of future development, the main limitation of ISF lies in the difficulty of achieving efficient and completely non-invasive access.

Saliva- and tear-based wearable sensors are much less widely applied than ISF-based wearable sensors, with most of them still in the stages of research and prototype development. Analytes in saliva also come from blood, with paracellular transport being the dominant transport pathway. However, unlike ISF, analyte concentrations in saliva are often significantly diluted (up to over 1000 times) compared to those in blood, which is one of its main limitations \cite{Heikenfeld2019,Mani2021}. In addition, the composition of saliva is less pure than that of ISF; in addition to metabolites from the human body, it also contains exogenous substances and microorganisms, which further increase the complexity of saliva analysis \cite{Mani2021}. The greatest advantage of saliva lies in its ease of non-invasive collection, which allows simple, rapid, and frequent sampling, and makes it particularly suitable for certain groups of users (such as babies) \cite{Heikenfeld2019,Min2021,Mani2021}. From an application perspective, the main problem with saliva sensors is the introduction of foreign objects into the oral cavity, which can cause discomfort during long-term use and may even lead to pathological issues such as inflammation \cite{Min2021}. Representative saliva-based sensing platforms include a pacifier-based biosensor for salivary glucose monitoring developed by García-Carmona \textit{et al}. \cite{GarciaCarmona2019}, a mouthguard-based glucose monitoring platform developed by Arakawa \textit{et al}. \cite{Arakawa2016}, and a tooth-integrated salivary sensor for Helicobacter pylori detection developed by Mannoor \textit{et al}. \cite{Mannoor2012}.

Tears are the least advantageous among the biofluids under discussion, primarily due to the difficulty in their acquisition \cite{Heikenfeld2019}. Tears are produced through the filtration of blood plasma and thus retain most of the smaller analytes while having a lower protein content. Compared to other non-invasive biofluids (i.e., saliva and sweat), the analyte composition of tears is simpler and more stable, which is mainly due to this filtration process \cite{Min2021,Li2022,Barmada2020}. Representative tear-based sensing platforms include a fluorescent scleral lens for tear electrolyte detection developed by Yetisen \textit{et al}. \cite{Yetisen2019}, a nose-bridge pad of eyeglasses integrated with a tear-sensing platform for the detection of alcohol, glucose, and vitamins developed by Sempionatto \textit{et al}. \cite{Sempionatto2019}, and a contact lens with an embedded tear glucose sensor developed by Yao \textit{et al}. \cite{Yao2011}.

Sweat is a highly promising target for wearable health monitoring. Compared with ISF, it offers the advantage of being completely non-invasive; and compared with saliva and tear, it can be obtained relatively easily without the need to introduce foreign objects into sensitive body areas, thereby minimizing interference with users' daily activities and making it suitable for long-term use. The typical form of wearable sweat sensors is flexible epidermal sensors that conform to the surface of the skin \cite{Min2021}. Unlike conventional rigid sensors, these sensors are better suited for application on soft, curving, and elastic skin, ensuring stable and intimate contact for reliable operation, while reducing discomfort during prolonged use. These advantages are largely attributed to advances in material science and microfabrication technologies \cite{Voldman1999}. The flexible sensors employ flexible substrates, primarily polymers such as PDMS, as well as flexible conductive materials, including graphene, carbon nanotubes (CNTs), and conductive polymers such as PEDOT:PSS. These materials enable the devices to meet electrochemical requirements while maintaining good mechanical properties and biocompatibility. In addition, advanced microfabrication techniques (e.g., photolithography and soft lithography) and printing techniques (e.g., screen printing, inkjet printing, and 3D printing) are widely applied and provide essential support for their miniaturization \cite{Han2017a,Huang2018,Costa2019}.

Wearable sweat sensing platforms are supported by well-established technologies for sweat acquisition and collection. For sweat acquisition, methods can be classified into two categories: natural or physiological perspiration and pharmacological-induced perspiration. Natural or physiological perspiration refers to the production of sweat through physiological responses to stimuli such as exercise, thermal stimulation, or psychological stress. Pharmacological-induced perspiration is mainly based on stimulating local sweat glands through chemical agents, which can be administered by epidermal application, iontophoresis, or injection to induce sweat secretion. These chemical agents are typically cholinergic drugs that activate muscarinic receptors in the sweat glands \cite{Bariya2018,Davis2024}.

For sweat collection, microfluidic systems are an increasingly reliable and widely adopted choice in modern wearable sensing platforms \cite{Heikenfeld2019,Min2021,Bariya2018,Davis2024,Chen2019,Choi2018}. Microfluidics is a technology for manipulating extremely small volumes of liquids ($10^{-9} \text{ to } 10^{-18}$) on the submillimeter scale ($\sim 50\,\mu\mathrm{m}$). At this scale, the forces dominating fluid behavior differ significantly from those on the macroscopic scale. For example, the relative influence of gravity on liquids is greatly diminished in comparison to surface and interfacial tension, as well as capillary forces. Moreover, while inertia often dominates in macroscopic flows rather than viscosity, the opposite is true at the microfluidic scale. The Reynolds number ($Re$) is introduced as a dimensionless quantity to characterize the ratio of inertial to viscous forces in a fluid, and it is proportional to the characteristic velocity and inversely proportional to the dynamic viscosity. In different ranges of $Re$, the flow regime of a fluid varies. When $Re \lesssim 2000$, which is the typical scenario in microfluidics, the fluid is mainly in laminar flow. In this case, the fluid streams run in parallel, with mixing occurring only through molecular diffusion, resulting in a low mixing efficiency. When $Re \gtrsim 2000$, which corresponds to most macroscopic scenarios, such as mixing milk and coffee or smoke and air, the primary flow regime begins to enter the transition region and changes to turbulence. In this case, the flow becomes irregular, with many eddies in the flow field, significantly accelerating mixing \cite{Whitesides2006,Sackmann2014}. In wearable sensing platforms, microfluidic systems are used to collect body fluid samples and guide them to specific reservoirs or detection sites. Microfluidics can drive sweat into the inlet of the microchannels through the joint effect of capillary forces and the natural pressure associated with perspiration, and subsequently guide it through the microchannels into designated chambers \cite{Chen2019}. Traditional sweat collection methods include the use of patches and film-based sweat storage pouches on the skin surface \cite{Kidwell2001,Brisson1991,Boysen1984}, but are rarely adopted in modern highly integrated wearable sweat monitoring platforms.

In the subsequent sections, this review first provides a detailed discussion of sweat biomarkers, including their classifications, physiological concentrations, partitioning mechanisms, and their significance in reflecting human physiological status. Next, this review examines the biosensors used in wearable sweat monitoring, classified according to their underlying detection mechanisms. Particular emphasis is placed on electrochemical sensors, the dominant modality in current wearable platforms, including their working principles and representative implementations in practical device designs. Finally, from a system-level perspective, this review analyzes the signal-processing strategies of integrated wearable sweat-monitoring platforms, covering common circuit architectures, signal transduction pathways, amplification and filtering schemes, calibration approaches, power management, and communication interfaces with user-end devices.

\section{Biomarkers in Sweat}
The biomarkers in sweat can be classified into the following classes: electrolytes, pH, metabolites, minerals, hormones, cytokines, proteins \& peptides, nutrients, and exogenous substances \cite{Min2023,Gao2023,Xu2021,Perez2022,Brothers2019,Bandodkar2019a}. Before discussing their physiological relevance, it is essential to first understand the mechanisms of analyte transport and partitioning in sweat. Since blood testing is the most established method for biofluid analysis in modern medicine, the link between blood analyte concentrations and physiological status has been studied much more extensively than other biofluids. Therefore, one of the most important ways to interpret the physiological implications of sweat biomarkers is to examine their correlations with blood levels. Alternatively, for biomarkers whose sweat concentrations do not directly correspond to blood, exploring their direct associations with human exertion offers another important avenue of interpretation \cite{Min2023,Gao2023}.

\begin{figure}[h]
    \centering
    \includegraphics[width=0.9\columnwidth]{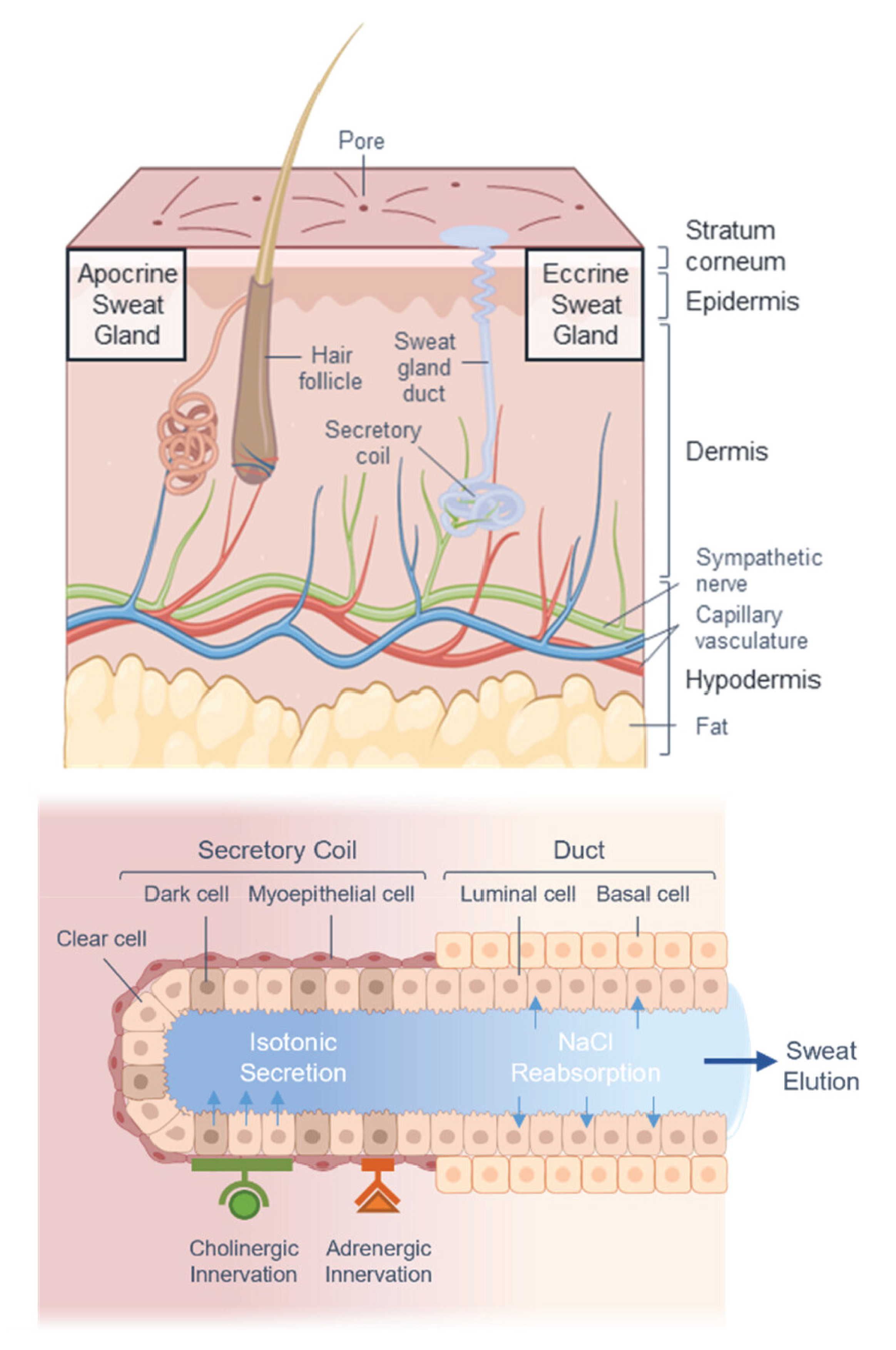}
    \caption{Structure of human skin and sweat glands. The upper illustration shows the structure of human skin, including apocrine and eccrine sweat glands embedded within the dermis. The lower illustration depicts the structure of an eccrine sweat gland, in which the secretory coil and the sweat duct constitute the two main components. The sweat gland lumen is enclosed by clear and dark cells at the secretory coil, and is lined by luminal ductal cells along the duct. Reproduced with permission from Ref.~\cite{Min2023}. \textcopyright~2023 American Chemical Society.}
    \label{fig:sweat_gland}
\end{figure}

The secretion pathways of different analytes in sweat are not completely identical. In addition, the composition of sweat obtained by natural secretion differs from that induced pharmacologically, and variations in the sweat flow rate can also result in significant changes in analyte composition \cite{Bariya2018,Davis2024,Min2023}. Therefore, each analyte needs to be analyzed individually. In general, one of the main pathways for analytes to enter sweat is blood–ISF–sweat. An eccrine sweat gland consists of three main parts: the secretory coil, the sweat duct, and the pore (see Figure~\ref{fig:sweat_gland}). The secretory coil is a convoluted tubular structure enclosed by sweat gland secretory cells, located in the dermis filled with interstitial fluid (ISF). The central space of the tubule, known as the sweat gland lumen, contains the primary sweat, which derives from the transport of water and dissolved constituents from the ISF. The sweat duct extends from the secretory coil, traverses the dermis, and terminates at the pore located on the skin surface. The primary sweat formed in the glandular lumen is isotonic relative to the ISF, blood, and the cytosol of secretory cells. However, as the fluid passes through the sweat duct, \ce{Na+} and \ce{Cl-} undergo reabsorption, which results in the final secreted sweat being hypotonic compared to plasma. Small, electroneutral, and lipophilic analytes (e.g., cortisol, ethanol) enter the ISF by crossing the plasma membrane of capillary endothelial cells via transcellular passive diffusion and subsequently cross that of sweat gland secretory cells in the same way to reach the glandular lumen. This transport process is relatively free, resulting in a strong correlation between their concentrations in blood and in sweat. However, for small, but charged and hydrophilic molecules (e.g., \ce{Na+}, \ce{Cl-}, \ce{K+}), since they cannot freely diffuse across the plasma membrane, their transports primarily relies on active mechanisms, with the \ce{Na+}-\ce{K+}-2\ce{Cl-} cotransport model \cite{Min2023} serving as a typical example. Large analytes (e.g., proteins, peptides), on the contrary, are hindered from the transcellular route due to their size. Instead, their transport mainly relies on the paracellular route through intercellular clefts between adjacent cells \cite{Heikenfeld2019,Min2023}. The following is a detailed discussion of each class of analytes.

\subsection{Electrolytes}
Electrolytes in sweat include mainly sodium, potassium, chloride \cite{Min2023,Gao2023,Xu2021,Perez2022,Brothers2019,Bandodkar2019a} and ammonium \cite{Min2023,Gao2023,Xu2021,Perez2022,Bandodkar2019a}. For other metal ions (e.g., \ce{Ca^2+}, \ce{Mg^2+}, \ce{Fe^2+}), since their sweat concentrations are relatively low, their importance in monitoring lies not primarily in the role of electrolytes, but rather in the context of minerals, which will be discussed later. \ce{Na+} and \ce{Cl-} are the most abundant ions in human sweat \cite{Min2023,Gao2023,Xu2021}, and serve as key regulators of sweat osmolality. During the formation of primary sweat, \ce{Na+} and \ce{Cl-} are actively pumped into the glandular lumen, increasing the osmolality of the luminal fluid and subsequently drives water influx along the osmotic gradient \cite{Matzeu2015}. \ce{Na+} is widely recognized as a key biomarker that reflects electrolyte balance and hydration status in the human body \cite{Min2023,Gao2023,Xu2021}, while \ce{Cl-} plays a pivotal role in the diagnosis and monitoring of cystic fibrosis \cite{Min2023,Gao2023,Xu2021,Bandodkar2019a}. The sodium concentration in sweat is related to multiple factors. Studies have shown that sweat sodium concentration is positively correlated with sweat rate. This is because, as sweat rate increases, both the secretion rate and the reabsorption rate of \ce{Na+} rise linearly, but the slope of the increase in secretion is markedly higher than that of reabsorption \cite{Buono2008}. In addition, evidence indicates that sweat sodium concentration is influenced by diet, with a high salt intake leading to a notable increase in sweat sodium levels. This suggests that perspiration may also serve as a mechanism to maintain sodium homeostasis in the human body \cite{Braconnier2018}. Moreover, sweat sodium concentration can also be associated with environmental adaptation. Studies have shown that after heat acclimatization, the sodium content in human sweat becomes relatively lower compared to the pre-acclimatization state \cite{Bates2008,Buono2007}.

\ce{K+} is another important electrolyte analyte in sweat. Blood potassium concentration is closely associated with muscle and nerve function, and reduced levels of blood potassium may cause muscle cramps \cite{Min2023,Gao2023,Xu2021}. Potassium in sweat originates from the blood \cite{Xu2021} and therefore exhibits a certain degree of correlation with blood potassium levels. In primary sweat, the concentration of \ce{K+} is isotonic with that of plasma; however, as the fluid passes through the sweat duct, the potassium concentration increases, resulting in a final potassium level of sweat that is slightly higher than that of plasma \cite{Sato1973,Buono2015}. A review has suggested that potassium concentration could serve as a predictor of muscle activity \cite{Xu2021}; however, the study cited \cite{Nemiroski2014} does not provide direct evidence supporting this relationship. Regarding the variation in sweat potassium under different conditions, some studies have reported that sweat potassium levels do not show a significant correlation with sweat rate \cite{Patterson2000,Schwartz1956}. Although some insights have already been gained, the partitioning mechanism of sweat potassium remains to be further elucidated, including the partitioning of \ce{K+} during the formation of primary sweat and the changes in its concentration as the fluid traverses the duct.

Studies have shown that \ce{NH4+} in sweat originates mainly from plasma \ce{NH3}.  \ce{NH3} enters the glandular lumen through passive transcellular diffusion, where it is protonated under acidic conditions. Because \ce{NH4+} is charged, its transmembrane transport is hindered and it becomes “trapped” in the lumen, resulting in total ammonia levels of sweat (\ce{NH3} and \ce{NH4+}) 20--50 times higher than in plasma. The concentration of sweat ammonium has been reported to be negatively correlated with sweat rate and pH \cite{Sato2005}. As a biomarker, blood ammonium is associated with liver diseases such as cirrhosis and acute liver failure, which lead to impaired hepatic clearance and consequently elevated blood ammonium levels. Furthermore, skeletal muscle releases \ce{NH4+} during exercise, suggesting that ammonium has the potential to help monitor exercise intensity \cite{Adeva2012}. Although current evidence supports the plasma-derived origin of sweat ammonium, the possibility of local production from sweat gland metabolism cannot be excluded and requires further investigation \cite{Sato2005}.

\subsection{pH}
A study has reported that under 5\% \ce{CO2} conditions, the pH of primary sweat ranges approximately 7.2--7.38, which is slightly lower than that of blood and does not vary with sweat rate. In contrast, the pH of the final secreted sweat falls within 5.0--7.0 and is influenced by the sweat rate. At low flow rates, the pH is around 5.0, whereas at high flow rates it increases to 6.5--7.0 \cite{Sato1989}. The pH of sweat directly influences the chemical forms of certain analytes. A typical example is the dynamic equilibrium between \ce{NH3} and \ce{NH4+}. Assuming that the total concentration of ammonia in sweat remains nearly constant, when the pH of sweat is one unit lower than that of plasma, the \ce{NH4+} content in sweat is approximately 10 times higher than in plasma; when the difference reaches two units, this value increases to approximately 100 times higher \cite{Sato2005}.

\subsection{Metabolites}
Metabolites in sweat include mainly glucose, lactate \cite{Min2023,Gao2023,Xu2021,Perez2022,Brothers2019,Bandodkar2019a}, ammonia, urea \cite{Min2023,Gao2023,Xu2021,Perez2022,Brothers2019}, uric acid \cite{Min2023,Gao2023,Xu2021,Perez2022}, and creatinine \cite{Min2023}. Continuous glucose monitoring (CGM) plays a critical role in the management of diabetes. Current CGM devices are mostly minimally invasive sensing platforms that target ISF, while sweat provides a promising route toward non-invasive glucose monitoring. As a relatively large and polar molecule, glucose enters sweat primarily through paracellular transport \cite{Jajack2018}. Experimental evidence demonstrates a strong correlation between sweat glucose and blood glucose concentrations. After excluding the interference of glucose from the stratum corneum and the ISF in the dermis, the measured sweat glucose concentration is approximately 1--2\% of that in the blood \cite{Moyer2012}. Another experimental study demonstrated that sweat glucose flux per unit of time is independent of sweat rate, which implies an inverse relationship between sweat glucose concentration and sweat rate \cite{Jajack2018}.

Regarding the partitioning mechanism of lactate in sweat, a widely accepted view is that it is not strongly correlated with plasma lactate but rather originates from local anaerobic metabolism within sweat gland cells \cite{Fellmann1983}. It is well known that during intense exercise, muscle cells break down carbohydrates through anaerobic glycolysis to rapidly obtain energy, thereby producing lactate. An early study reported that while blood lactate levels increased significantly after exercise, sweat lactate levels were not markedly affected \cite{Weiner1952}. In addition, lactate concentration decreases as sweat rate increases and drops further as sweating continues, which can be attributed to a dilution effect \cite{Fellmann1983,Weiner1952}. As a biomarker, some reviews suggest that lactate has the potential to indicate muscle exertion, muscle fatigue \cite{Min2023}, tissue viability, and stress ischemia \cite{Gao2023}.

As nitrogen-containing compounds, ammonia, urea, uric acid, and creatinine could serve as indicators of renal function \cite{Min2023}. Ammonia, as another form of ammonium, maintains a dynamic equilibrium with it in sweat. Since its transport mechanism and biomarker significance have already been discussed in the context of ammonium, no further elaboration is provided here. Some studies have shown that sweat urea originates primarily from plasma and enters sweat through transcellular passive diffusion \cite{Sato1989,Komives1966}. The sweat-to-plasma ratio is close to 1:1 and remains stable across a wide range of plasma concentrations and sweat rates. This indicates that metabolism within sweat glands is not the major source of sweat urea \cite{Komives1966}. Uric acid is well known as an important indicator for the diagnosis of gout \cite{Gao2023}. Unlike urea, the concentration of uric acid in sweat is very low (approximately 6.3\% of that in serum \cite{Huang2002}) and does not show a significant correlation with serum uric acid levels \cite{Huang2002,AlTamer1997}. At the same time, both of these studies suggested that uric acid in sweat may originate from a "leakage" of serum \cite{Huang2002,AlTamer1997}. Given the relatively large size and strong polarity of uric acid, it is hypothesized that it may enter sweat through paracellular transport, while the specific partitioning mechanism remains to be further investigated. Regarding creatinine, some studies found a similar situation to uric acid \cite{Huang2002}, while its exact partitioning mechanism is unclear.

\subsection{Minerals}
Mineral analytes in sweat mainly include calcium \cite{Min2023,Gao2023,Xu2021,Perez2022,Brothers2019,Bandodkar2019a}, magnesium \cite{Min2023,Perez2022}, zinc \cite{Min2023,Gao2023,Xu2021,Bandodkar2019a}, copper \cite{Gao2023,Xu2021}, and iron \cite{Min2023,Gao2023}. These elements are essential macrominerals or trace elements for the human body, whereas toxic heavy metals (e.g., \ce{Pb^2+}, \ce{Hg^2+}) will be discussed in the section on Exogenous substances. Calcium, one of the macrominerals, is the fifth most abundant element in the human body. In body fluids, it exists in several forms, including protein-bound, ligand-complexed, crystalline salts, and ionized (\ce{Ca^2+}), with the ionized form being most directly associated with physiological processes. \ce{Ca^2+} accounts for approximately half of the total calcium in plasma and an even higher proportion in other body fluids (e.g., urine and saliva). As a biomarker, \ce{Ca^2+} levels are of clinical significance in the diagnosis and monitoring of hyperparathyroidism, renal failure, acute pancreatitis, and multiple myeloma \cite{Robertson1981}. Several studies have investigated calcium concentrations in sweat \cite{Ely2011,Montain2007,Chinevere2008,Baker2011}, with findings indicating that sweat total calcium concentration is markedly lower than that in plasma (approximately one-sixth to one-seventh \cite{Ely2011}), remains relatively stable during continuous sweating \cite{Montain2007}, but decreases after heat acclimatization \cite{Montain2007,Chinevere2008}. Magnesium, another macromineral in the human body, exhibits a pattern in sweat similar to calcium. Sweat total magnesium concentration is likewise lower than that in plasma (approximately one-fourteenth \cite{Ely2011}), shows a slight declining trend during continuous sweating \cite{Montain2007}, and decreases notably after heat acclimatization \cite{Chinevere2008}. Although many studies and reviews have mentioned magnesium as a common mineral analyte in sweat and it indeed plays an important role in the human body, its overall clinical significance as a biomarker remains limited and requires further investigation \cite{Witkowski2011}.

Zinc, copper, and iron are essential trace elements in the human body, and both their deficiency and excess can lead to health problems. Zinc deficiency may lead to immune system impairment, delayed wound healing, and impaired senses of taste and smell, whereas excessive intake may result in reduced immune function \cite{Fraga2005}. The average concentration of total zinc in sweat is approximately one-third to one-fourth of that in plasma \cite{Ely2011}, and it decreases during continuous sweating. This suggests the presence of zinc conservation, meaning that the body may employ certain mechanisms to reduce zinc loss through sweat \cite{Montain2007}. Sweat total zinc concentration also shows a decreasing trend after heat acclimatization, although the change did not reach statistical significance \cite{Chinevere2008}. Copper deficiency is rare in humans (but may be induced by excessive zinc intake). When it occurs, it can lead to normocytic, hypochromic anemia, leucopenia, and neutropenia, whereas acute copper intoxication may cause gastrointestinal effects \cite{Fraga2005}. The total copper concentration in sweat is much lower than that in plasma, accounting for only about one-twentieth \cite{Ely2011}. Its variation under different external conditions, such as continuous sweating and heat acclimatization, is similar to that of magnesium \cite{Montain2007,Chinevere2008}. Iron is the most abundant trace element in the human body, with deficiency leading to anemia and excess causing severe organ damage \cite{Fraga2005}. The total iron concentration in sweat is negligible compared to plasma \cite{Ely2011}, remains largely unchanged during continuous sweating \cite{Montain2007}, and decreases after heat acclimatization, although the reduction does not reach statistical significance \cite{Chinevere2008}.

Regarding the partitioning mechanisms of these minerals in sweat, available evidence remains scarce. A review has suggested that they may enter sweat via paracellular pathways, likely due to their small size, charge, and hydrophilic nature. It is also important to note that, similar to calcium, the other minerals discussed above also exist in body fluids to a considerable extent in non-ionized forms, particularly iron, which accounts for up to 95\% \cite{Min2023}.

\subsection{Hormones}
Cortisol, a steroid hormone, is the most frequently reported hormonal analyte in sweat and is closely associated with the regulation of the human stress response \cite{Min2023,Gao2023,Xu2021,Perez2022,Brothers2019,Bandodkar2019a}. During stress, it mediates signaling that promotes fat and protein catabolism, as well as glucose synthesis, thus rapidly providing energy to cope with threats. At the same time, it inhibits the transcription of pro-inflammatory genes, thereby reducing the synthesis and secretion of pro-inflammatory cytokines and preventing tissue damage caused by excessive immune responses \cite{Smyth2013,Lee2015,Russell2019}. Regarding the partitioning mechanism of cortisol into sweat, as a lipophilic molecule, it could in principle diffuse relatively freely across the plasma membrane. However, because free cortisol in the blood is limited, while the majority is bound to carrier proteins, only a very small fraction can actually diffuse into sweat, resulting in a much lower total cortisol concentration in the sweat than in serum \cite{Min2023,Perez2022,Brothers2019}. Nevertheless, some studies have shown that trends in sweat cortisol diurnal variation are correlated with those of serum and salivary cortisol \cite{Wang2022a,TorrenteRodriguez2020}, although detection standards need to be further standardized \cite{Pearlmutter2020}. Beyond cortisol, dehydroepiandrosterone (DHEA), another steroid hormone associated with circadian rhythm, has also been shown to be a promising biomarker for sweat-based chronobiology monitoring \cite{Upasham2020a,Upasham2020b}.

Neuropeptides, although not entirely equivalent to classical hormones, are often discussed as hormone-like molecules in the context of biomarkers. Neuropeptide Y (NPY) is one of the commonly detected neuropeptide analytes in sweat \cite{Min2023,Gao2023,Perez2022}. NPY is a 36 amino acid polypeptide that is closely associated with the regulation of brain activity, stress response, digestion, blood pressure, heart rate, metabolism, and immunological function \cite{Zhang2021}. A study reported that plasma and sweat NPY levels were significantly elevated in patients with major depressive disorder (MDD) compared to healthy controls, and a strong positive correlation was observed between them \cite{Cizza2008}.

\subsection{Proteins \& Peptides}
A study reported that the protein composition of human sweat is highly imbalanced. The researchers have detected 95 proteins in sweat, among which the five most abundant accounted for 91\% of the total content, namely dermcidin (46\%), clusterin (17\%), apolipoprotein D (15\%), prolactin-inducible protein (PIP) (8\%), and serum albumin (6\%). All of these proteins are closely associated with skin defense and protection through different mechanisms such as antimicrobial activity, anti-inflammatory regulation, or antioxidative effects \cite{Csosz2015}. In addition to these, some trace proteins in sweat have also attracted interest, particularly cytokines and C-reactive protein (CRP). Cytokines are small proteins or peptides secreted by cells that serve mainly as signaling molecules to transmit information between cells. They play an important role in the human immune system and inflammatory response. Although the exact partitioning mechanism of cytokines in sweat remains unclear \cite{Hladek2018}, a study has shown that the levels of several cytokines in sweat, including IL-1$\alpha$, IL-1$\beta$, IL-6, TNF-$\alpha$, IL-8, and TGF-$\beta$, show significant correlations with their plasma levels \cite{MarquesDeak2006}. Furthermore, sweat levels of three pro-inflammatory cytokines, TNF-$\alpha$, IL-6, and IL-10, have been found to be higher in older individuals than in younger ones \cite{Hladek2018}, which is consistent with established patterns of cytokine levels in blood. CRP is another acute phase protein that is closely associated with inflammatory responses. A study first reported the presence of CRP in sweat and quantitatively measured its concentration, demonstrating the feasibility of accurately detecting sweat CRP using a wearable sweat sensing platform. The experimental results showed that CRP levels in sweat were much lower than normal levels in blood (at the $10^{-5}$ level) \cite{Jagannath2020}. Meanwhile, the relationship between sweat CRP levels and the gold standard, as well as the partitioning mechanism, remains largely unexplored and required further investigation.

In general, the partitioning mechanisms of sweat proteins are complex. There is evidence that some proteins originate from local endogenous secretion of sweat glands, while others may be derived from the diffusion of plasma components. However, most of these mechanisms remain unclear and require further investigation \cite{Csosz2015,Hladek2018}.

\subsection{Nutrients}
Sweat is also known to contain relatively smaller molecular nutrients compared to proteins and peptides, primarily vitamins \cite{Min2023,Bandodkar2019a} and amino acids \cite{Min2023,Gao2023,Xu2021,Perez2022,Bandodkar2019a}. A study has reported that vitamin C intake results in elevated vitamin C levels in sweat, and sweat vitamin C shows a strong positive correlation with blood vitamin C \cite{Zhao2020}. This highlights the potential of sweat as a non-invasive alternative for vitamin C monitoring. In addition, dehydroascorbic acid, various B-complex vitamins (e.g., thiamine, riboflavin, nicotinic acid), and some vitamin-like compounds (e.g., inositol, choline) have also been detected in sweat \cite{Harvey2010}.

A study tried to detect 26 amino acids in human sweat (three of which were not detected), with serine, glycine, alanine, citrulline, and threonine being the most abundant. In particular, methionine, cysteine, and tryptophan, which are very common in plasma, were only present in trace amounts or were undetectable in sweat \cite{Mark2012}. An early study has shown that the amino acid levels in sweat were generally equal to or higher than those in plasma. Although fluctuations were observed between different individuals and across repeated experiments in the same individual, the overall relationship remained stable. The study has also shown, after oral ingestion of a specific amino acid, the levels of the corresponding amino acids in plasma and urine increased significantly, whereas those in sweat were not significantly affected. This suggests that amino acids in sweat are not derived primarily from plasma diffusion but are more likely to be secreted endogenously by sweat glands \cite{Hier1946}. Compared to blood, the amino acid composition of sweat is more closely reminiscent of that of the epidermal barrier protein filaggrin. Therefore, the partition of amino acids into sweat is more likely the result of equilibration or exchange with filaggrin degradation products in the stratum corneum during sweat secretion, which are the main constituents of the natural moisturizing factor (NMF) of the skin \cite{Mark2012,Dunstan2016}.

\subsection{Exogenous substances}
The exogenous substances present in sweat include alcohols, drugs, and heavy metals \cite{Min2023,Gao2023,Xu2021,Perez2022,Brothers2019,Bandodkar2019a}, while perspiration serves as an important pathway for their excretion. Ethanol is one of the most frequently discussed analytes in sweat monitoring. Approximately 90\% of the ethanol ingested is sequentially metabolized into acetaldehyde, acetate, and acetyl coenzyme A (CoA) through the catalysis of multiple enzymes, while a portion of the remaining ethanol is eliminated directly through breath, urine, and sweat \cite{Min2023}. As everyone knows, excessive alcohol consumption can lead to violence, traffic accidents, and various health issues; therefore, convenient and efficient alcohol detection is of great significance. Currently, existing alcohol detection devices include platforms based on blood, urine, saliva, and breath. However, most of these methods are limited by their dependence on complex instrumentation or invasive biofluid sampling. The most common breath alcohol analyzer, although convenient, is easily affected by environmental factors leading to limited accuracy \cite{Kim2016}. Sweat-based alcohol detection offers a promising alternative. Studies have shown that ethanol concentrations in sweat and blood can rapidly reach equilibrium, with an almost 1:1 relationship. This is because ethanol, being a small molecule with moderate lipid solubility, can easily penetrate the plasma membrane of cells and thus easily diffuse through sweat gland epithelial cells into the glandular lumen \cite{Buono1999}. This property allows sweat ethanol to accurately reflect blood alcohol levels. In recent years, sweat-based alcohol sensing platforms have been developed and demonstrated to be functionally feasible \cite{Kim2016,Hauke2018}.

Caffeine and levodopa (L-dopa) are two drugs that have been frequently discussed in the context of sweat-based monitoring. Caffeine monitoring plays an important role in anti-doping surveillance. A study has shown that caffeine levels in sweat increase significantly with caffeine intake and exhibit strong positive correlations with plasma and urinary caffeine concentrations. This indicates that sweat caffeine levels can reflect the overall body caffeine load and have great potential as an alternative target biofluid for caffeine monitoring \cite{Kovacs1998}. Furthermore, a recent study developed a wearable sweat-based caffeine monitoring platform, which successfully achieved real-time detection of caffeine concentration in sweat and validated the correlation between sweat, blood, and urine caffeine levels \cite{Tai2018}. Levodopa is the standard medication for the treatment of Parkinson’s disease, and monitoring L-dopa levels plays a crucial role in optimizing drug dosage. A study developed a wearable sweat-based L-dopa monitoring platform that enables non-invasive, real-time detection of L-dopa and verified that the temporal variation of L-dopa concentration in sweat exhibits a similar trend to that in plasma \cite{Tai2019}.

Many heavy metals do not have any physiological benefit for the human body and can cause severe damage through various mechanisms. A systematic review summarized studies on the metabolism of four toxic metals--arsenic, cadmium, lead, and mercury--in sweat, and found that perspiration plays a significant role in their excretion, with the amount eliminated through sweat in some cases being comparable to or even exceeding that in urine \cite{Sears2012}. This highlights the potential importance of perspiration as an alternative pathway for heavy metal detoxification.

\section{Biosensors for wearable sweat monitoring}
Wearable health monitoring devices targeting biofluids rely on biosensors to transduce biochemical information (e.g., concentration) into measurable and quantifiable signals (e.g., electric signal, optical signal). An ideal biosensor should exhibit a fast response time, high sensitivity and selectivity, low limit of detection / quantification, minimal hysteresis, long-term stability, and robustness under varying operational conditions. In the context of wearable sweat monitoring, additional design considerations include lightweight construction, energy efficiency, and good biocompatibility. However, simultaneously achieving all of these desirable parameters in practical applications is extremely challenging and often nearly impossible. Therefore, it is essential to identify the most critical performance parameters during the device design stage and make reasonable trade-offs among the less crucial \cite{Min2023}.

Some types of sensors possess these desired characteristics in specific sweat-sensing applications. Based on their detection mechanisms, the sensors employed in wearable sweat monitoring platforms can be broadly classified into electrochemical sensors, optical sensors, physical sensors, and biorecognition-based sensors. The selection of a particular sensor is primarily determined by the intrinsic properties of the target analytes. In the following sections, the working principles of different detection mechanisms are described, and the applications of various sensors in practical wearable sweat-sensing designs are reviewed and discussed.

\subsection{Electrochemical sensors}
Electrochemical detection is the dominant analytical approach in wearable and non-invasive health monitoring, forming the core of nearly all designs that aim to analyze biological samples at the molecular level. This prominence arises from its intrinsic ability to transduce variations in analyte concentration into corresponding electrical responses in real time, while functionalized electrodes enable high sensitivity and selectivity toward specific targets. In sweat analysis, electrochemical detection techniques include potentiometry, amperometry, voltammetry, and emerging field effect transistor (FET)-based methods \cite{Min2023, Bandodkar2019a}. The selection among these techniques depends mainly on the molecular characteristics and concentration range of the target analytes \cite{Gao2023}.

\subsubsection{Potentiometric Sensors}
\begin{figure}[h]
    \centering
    \includegraphics[width=0.9\columnwidth]{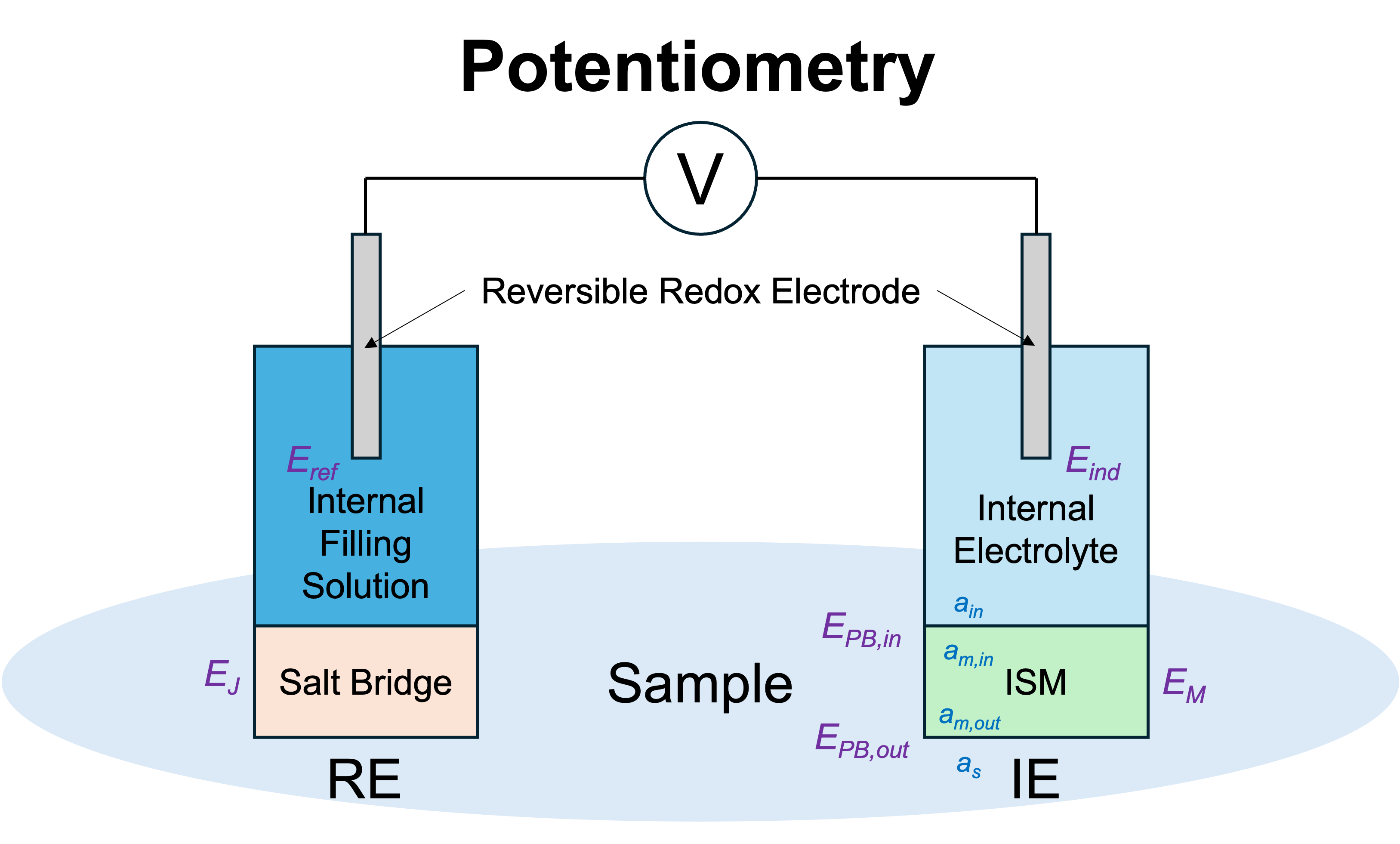}
    \caption{Schematic illustration of a potentiometric two-electrode system, where the open-circuit potential between the IE and the RE is measured. IE: indicator electrode; RE: reference electrode.}
    \label{fig:pot_schematic}
\end{figure}

Potentiometry \cite{Clarke2020} is an electrochemical analytical method that measures the open circuit potential between two electrodes to reflect the analyte concentration. It is commonly used for the detection of ionic analytes, whose activity can be correlated with the measured potential difference. A typical potentiometric system consists of a reference electrode that provides a stable and reproducible potential, and an indicator electrode whose potential varies with the activity of the target analyte in the sample solution (see Figure~\ref{fig:pot_schematic}). A voltmeter with a very high input impedance is connected between the two electrodes to measure the potential difference, ensuring that the current through the cell remains negligible and that electrochemical equilibrium at both electrodes is maintained. The relationship between the electrode potential and the analyte activity obeys the Nernst equation:
\begin{equation}
E = E^{\circ} - \frac{RT}{nF} \ln Q
\end{equation}
where $E$ is the electrode potential, $E^{\circ}$ is the standard electrode potential, $R$ is the gas constant ($8.314~\mathrm{J{\cdot}mol^{-1}{\cdot}K^{-1}}$), $T$ is the temperature ($\mathrm{K}$), $n$ is the number of electrons transferred in the electrode reaction, $F$ is the Faraday constant ($96\,485~\mathrm{C{\cdot}mol^{-1}}$), and $Q$ is the reaction quotient.

In a potentiometric system, the reference electrode is typically a \ce{Ag}/\ce{AgCl} electrode (can also be a \ce{Hg}/\ce{Hg2Cl2} electrode, etc.) immersed in a concentrated solution of \ce{KCl}. A reversible redox equilibrium occurs at the interface between the electrode and the electrolyte:
\begin{equation}
\mathrm{AgCl(s) + e^- \rightleftharpoons Ag(s) + Cl^-(aq)}
\end{equation}
Because both \ce{AgCl} and \ce{Ag} are solid phases with constant activities of 1, the reaction quotient depends solely on the activity of \ce{Cl^-}. The corresponding Nernst equation can be expressed as:
\begin{equation}
E_{\ce{Ag/AgCl}} = E^{\circ}_{\ce{Ag/AgCl}} - \frac{RT}{F} \ln a_{\ce{Cl^-}}
\end{equation}
Here, $E_{\ce{Ag/AgCl}}$ is the potential of the \ce{Ag}/\ce{AgCl} electrode, $E^{\circ}_{\ce{Ag/AgCl}}$ represents its standard potential, which is approximately $+0.222~\mathrm{V}$ versus the standard hydrogen electrode (SHE) at $a_{\ce{Cl^-}} = 1$. The high concentration solution \ce{KCl} helps maintain constant chloride ion activity in the system, and when the solution is saturated, the electrode potential is approximately $+0.197~\mathrm{V}$. The internal electrolyte of the reference electrode is in contact with the sample solution through a porous membrane, forming a salt-bridge interface that ensures the closure of the electrochemical circuit. The porous membrane, with its small pore size, reduces the differences in ionic mobility at the interface, and thus minimizes and stabilizes the liquid junction potential.

The indicator electrode is typically an ion-selective electrode (ISE) \cite{Bakker1997}. In addition to an internal \ce{Ag}/\ce{AgCl} substrate immersed in an internal electrolyte, which serves as the ion-to-electron transducer, the ISE is separated from the sample solution by an ion-selective membrane (ISM). The ISM enables the electrode to respond selectively to a specific ion in the sample while excluding others, which is composed of several key components including a polymer matrix with a plasticizer, an ionophore, and an ion-exchanger. The polymer matrix, typically poly(vinyl chloride) (PVC), provides the structural framework and enables the mechanical stability of the membrane. It is electrically insulating and hydrophobic. Because PVC is intrinsically rigid, a plasticizer is added to modify its physical properties, rendering the membrane a viscous, water-immiscible liquid. This allows the ionophore and ion-exchanger to be uniformly distributed and mobile within the membrane phase. The ionophore is the key component that imparts ion selectivity to the ISM. It is a lipophilic organic molecule capable of forming specific and reversible complexes with target ions. Because the membrane is hydrophobic, only ions that can form stable complexes with the ionophore can partition into and permeate the membrane, whereas other hydrated ions in the aqueous phase are effectively excluded. The ion-exchanger also plays an essential role in the transmembrane transport of the target ion. In cation-selective membranes, it provides lipophilic anionic sites that balance the charge of incoming cations, thereby maintaining overall electroneutrality within the membrane. Without such ion-exchanger sites, the influx of ions would immediately generate a large electric field inside the membrane, preventing further ion transfer.

The ISM forms two interfaces with the aqueous phases: one between the outer side of the ISM and the sample solution, and the other between the inner side of the ISM and the internal electrolyte. At each interface, a phase-boundary potential is established. This potential arises from the equilibrium of ionic chemical potentials across the membrane-water interface. Since non-target ions cannot permeate the membrane and thus cannot establish interfacial equilibrium, they make no contribution to the interfacial potential under ideal conditions. The relationship between the interfacial potential and the ionic activity is described by the Nernst equation:
\begin{align}
E_{\mathrm{PB,out}}
&= E_{\mathrm{out}}^{\circ}
- \frac{RT}{zF}\ln\!\left(\frac{a_{\mathrm{m,out}}}{a_{\mathrm{s}}}\right) \\
E_{\mathrm{PB,in}}
&= E_{\mathrm{in}}^{\circ}
- \frac{RT}{zF}\ln\!\left(\frac{a_{\mathrm{m,in}}}{a_{\mathrm{in}}}\right)
\end{align}
where $E_{\mathrm{PB,out}}$ and $E_{\mathrm{PB,in}}$ represent the phase-boundary potentials at the outer and inner interfaces, respectively; $E_{\mathrm{out}}^{\circ}$ and $E_{\mathrm{in}}^{\circ}$ are constant terms specific to each interface; $z$ is the charge number of the target ion; $a_{\mathrm{s}}$ and $a_{\mathrm{in}}$ denote the activities of the target ion in the sample solution and the internal electrolyte, respectively; and $a_{\mathrm{m,out}}$ and $a_{\mathrm{m,in}}$ are the activities of the target ion in the membrane phase near the outer and inner interfaces, respectively. The total membrane potential is made up of the two interfacial potentials and the diffusion potential within the membrane:
\begin{equation}
E_{\mathrm{M}}
= E_{\mathrm{PB,out}}
- E_{\mathrm{PB,in}}
+ E_{\mathrm{diff}}
\end{equation}
Here, $E_{\mathrm{M}}$ represents the total membrane potential, and $E_{\mathrm{diff}}$ represents the diffusion potential. When the activity gradient of the ions within the membrane is zero, $E_{\mathrm{diff}} = 0$. And if the ionic activities near the outer and inner sides of the membrane phase are equal ($a_{\mathrm{m,out}} = a_{\mathrm{m,in}}$), the expression for $E_{\mathrm{M}}$ can be simplified as:
\begin{equation}
E_{\mathrm{M}}
= E_{\mathrm{M}}^{\circ}
+ \frac{RT}{zF}\ln\!\left(\frac{a_{\mathrm{s}}}{a_{\mathrm{in}}}\right)
\end{equation}
where $E_{\mathrm{M}}^{\circ}$ is a constant term that combines all fixed contributions. This relationship holds under the condition that the ionophore and ion-exchangers are uniformly distributed within the ISM, which represents the typical operational state of the membrane. It can be seen that the membrane potential depends only on the activities of the target ion in the sample solution and in the internal electrolyte. In practice, the concentration of the internal electrolyte is usually kept constant; therefore, the activity of the target ion in the sample solution can be directly reflected by the membrane potential.

For the entire system, the potential measured by the voltmeter is the potential difference between the indication and reference electrodes, corresponding to the difference in the local electric potentials at the metal contacts of the two electrodes:
\begin{equation}
E_{\mathrm{meas}}
= \phi_{\mathrm{ind}} - \phi_{\mathrm{ref}}
\end{equation}
The local electric potential at each metal contact can be expressed as the sum of the boundary potentials and the membrane potential along the electrochemical path. Accordingly, the measured potential can be written as follows:
\begin{equation}
E_{\mathrm{meas}}
= (E_{\mathrm{M}}
+ E_{\mathrm{ind}})
- (E_{\mathrm{ref}}
+ E_{\mathrm{J}})
\end{equation}
where $E_{\mathrm{ind}}$ and $E_{\mathrm{ref}}$ represent the boundary potentials established at the interfaces between each metal electrode and its internal electrolyte, and $E_{\mathrm{J}}$ is the liquid-junction potential between the reference electrode and the sample solution.
In practical operation, $E_{\mathrm{ind}}$, $E_{\mathrm{ref}}$, and $E_{\mathrm{J}}$ are constant for a given setup, so $E_{\mathrm{meas}}$ depends solely on $E_{\mathrm{M}}$, which directly reflects the activity of the target ion in the sample:
\begin{equation}
E_{\mathrm{meas}}
= E_{\mathrm{meas}}^{\circ}
+ \frac{RT}{zF}\ln a_{\mathrm{s}}
\end{equation}
where $E_{\mathrm{meas}}^{\circ}$ collects all constant terms. This builds up a detection system with the activity of the target ion as its input and the potential measured by the voltmeter as its output.

Conventional liquid-contact ISEs rely on the internal filling electrolyte between the ISM and the metal substrate as the ion-to-electron transducer. However, such ISEs are sensitive to the evaporation of the internal electrolyte as well as to variations in temperature and pressure \cite{Hu2016}. The osmotic pressure that results from the concentration difference between the sample and the internal electrolyte can further introduce the risks of internal water leakage and ISM delamination \cite{Lindner2008}. In addition, the liquid-contact configuration is unfavorable for electrode miniaturization \cite{Min2023,Hu2016}. These characteristics make conventional liquid-contact ISEs unsuitable for wearable applications. One solution is the use of solid-contact ISEs. Unlike liquid-contact ISEs, which rely on the reversible redox reaction occurring at the \ce{Ag}/\ce{AgCl} electrode–internal electrolyte interface to achieve ion-to-electron transduction, solid-contact ISEs employ different mechanisms that utilize redox capacitance or double layer capacitance \cite{Hu2016}.

Conventional capacitors store charge by accumulating positive and negative charges on two electrodes under an external electric field, whereas redox capacitance adopts a different mechanism that stores charge through reversible redox reactions, in which electrons are retained in the reduced state of the redox-active species. Solid-contact ISEs based on redox capacitance employ conducting polymers doped with anions (when the target ion is a cation) as the solid-contact layer material, commonly including polypyrrole (PPY) \cite{Cadogan1992}, poly(3-octylthiophene) (POT) \cite{Bobacka1994}, polyaniline (PANI) \cite{Bobacka1995}, and poly(3,4-ethylenedioxythiophene) (PEDOT) \cite{Bobacka1999}, among which PANI and PEDOT are most widely used in wearable applications \cite{Min2023}. The doped conducting polymer exhibits both electronic and ionic conductivity. When the target ion enters the solid-contact layer from the ISM, an equivalent amount of electron transfer occurs to maintain charge neutrality, leading to the reduction of the conducting polymer, i.e., the undoping process. The released anions form ion pairs with the incoming target ions within the solid-contact layer. Correspondingly, the ion exchanger (lipophilic anionic sites) within the ISM can also undergo a reversible redox interaction with the undoped conducting polymer, restoring the doped state. The relative contributions of these two processes are determined by the mobilities of the target ions and the anions serving as ion exchanger \cite{Hu2016}. The potential difference at the ISM/solid-contact interface depends on the distribution equilibrium of the target ion across the two phases. Under ideal conditions, the solid-contact layer forms an ohmic contact with the metallic substrate, in which the electronic chemical potentials are continuous and the interfacial potential difference is negligible. This situation slightly differs from that of liquid-contact ISEs, where the reversible redox reaction at the internal electrolyte/metal substrate interface gives rise to a phase boundary potential.

\begin{figure*}[tp]
    \centering
    \includegraphics[width=0.9\textwidth]{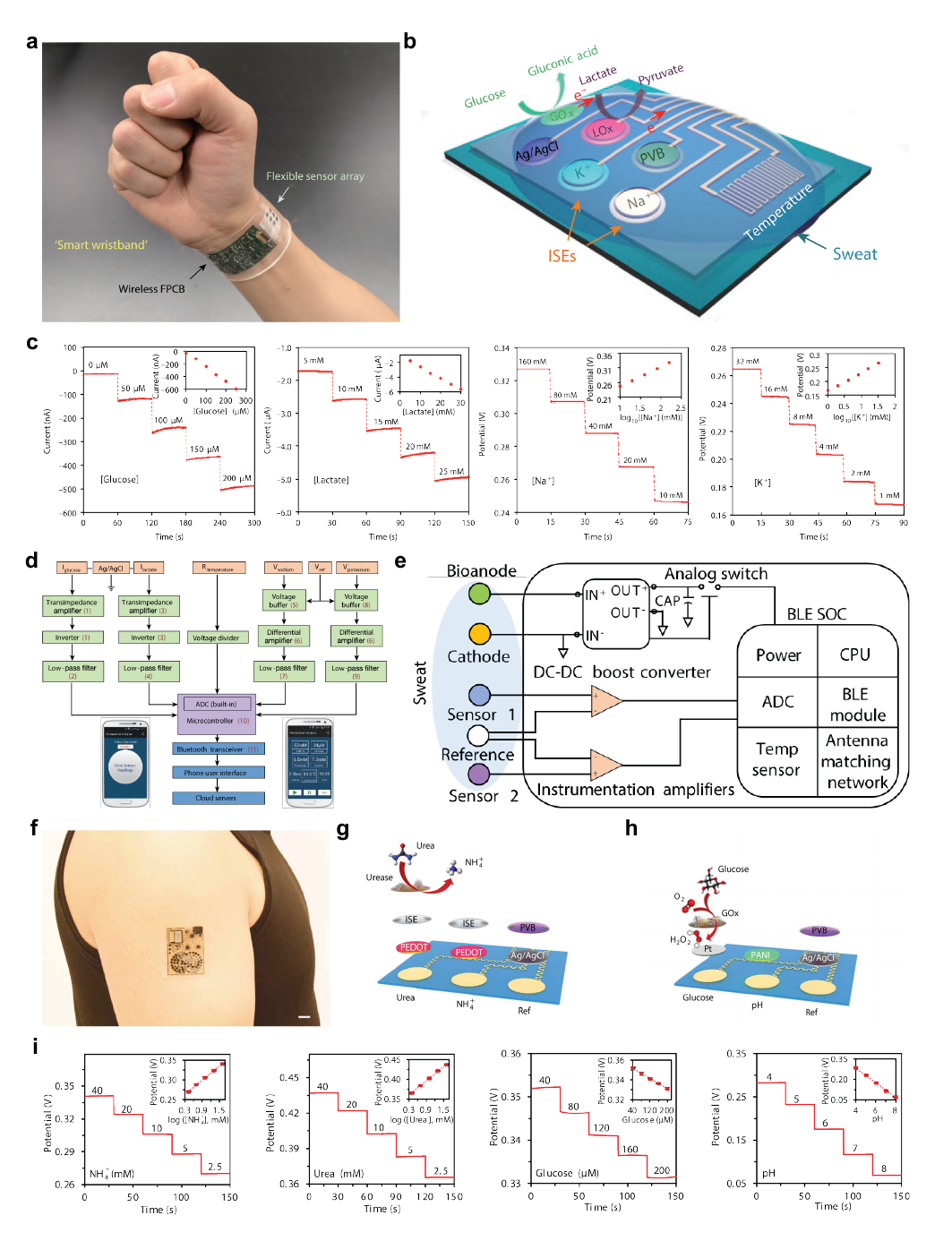}
    \caption{A fully integrated sensor array (FISA) (a–d) and a perspiration-powered electronic skin (PPES) (e–i) for wearable sweat analysis. (a) Photograph of the FISA worn on the wrist. (b) Schematic illustration of the multi-analyte sensor array. (c) Sensor responses, from left to right: the chronoamperometric responses of the glucose and lactate sensors, followed by the open-circuit potential responses of the sodium and potassium sensors. (d) Block diagram showing the system-level signal flow of the FISA. ADC: analog-to-digital converter. Reproduced from Ref. \cite{Gao2016}. (e) Schematic diagram of the PPES system. BLE: Bluetooth Low Energy; CPU: central processing unit; SoC: system-on-chip. (f) Photograph of the PPES worn on the arm. (g) Schematic illustration of the sensor array for \ce{NH4+} and urea detection. (h) Schematic illustration of the sensor array for pH and glucose detection. (i) Sensor responses, from left to right: the open-circuit potential responses of the \ce{NH4+}, urea, glucose, and pH sensors. Reproduced from Ref. \cite{Yu2020}.}
    \label{fig:pot_designs}
\end{figure*}

Solid-contact ISEs with double-layer capacitance rely on a direct capacitive coupling formed at the ISM/solid-contact layer interface to achieve ion-to-electron transduction. The solid-contact layer of this type of electrode possesses electronic but not ionic conductivity; as a result, the target ions in the ISM are unable to diffuse into the solid-contact phase. The target ions accumulate on the ISM side of the ISM/solid-contact interface, while an equivalent amount of electrons are induced and accumulate on the solid-contact side, forming an electrical double layer. In this case, the phase boundary potential is determined by the amount of charge stored in the electrical double layer, directly establishing a correlation with the activity of the target ion. The performance of the double-layer capacitance is closely related to the interfacial surface area where the electrical double layer is formed. A larger contact area facilitates higher capacitance and thus improves potential stability. Therefore, this type of ISE often uses nanomaterials as the solid-contact layer, which can enlarge the interfacial surface area while maintaining the same projected area. These typically include carbon-based nanomaterials such as three-dimensionally ordered macroporous (3DOM) carbon \cite{Lai2007}, carbon nanotubes \cite{Crespo2008}, fullerene (\ce{C_60}) \cite{Fouskaki2008,Li2015}, graphene \cite{Ping2011,Hernandez2012,Li2012,Miller2013}, colloid-imprinted mesoporous (CIM) carbon \cite{Hu2014}, and porous carbon spheres \cite{Ye2015}; as well as metallic nanomaterials such as metallic nanoparticles \cite{PaczosaBator2012,PaczosaBator2013,Zeng2020} and metal–organic frameworks \cite{Mendecki2018}.

\begin{figure}[h]
    \centering
    \includegraphics[width=0.9\columnwidth]{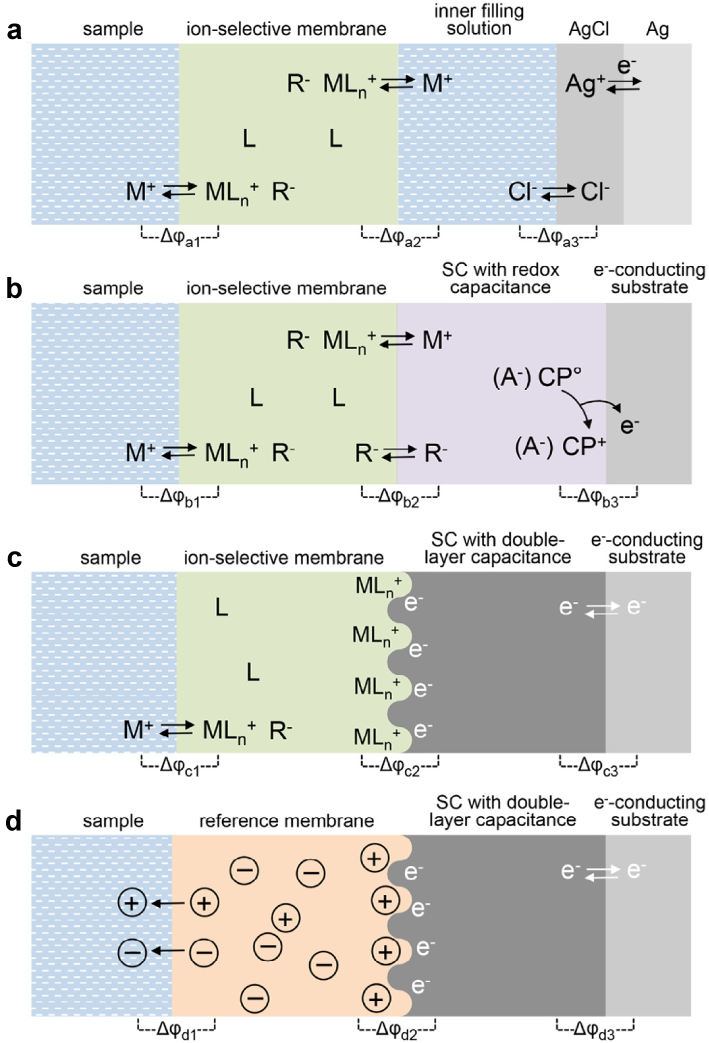}
    \caption{Schematic illustrations of liquid-contact ISEs, solid-contact ISEs based on redox / double-layer capacitance, and solid-contact reference electrodes; \ce{M+}: target ion, \ce{L}: ionophore, \ce{R-}: ion-exchanger providing anionic sites. (a) Liquid-contact ISE; (b) solid-contact ISE with dominant redox capacitance; (c) solid-contact ISE with high double-layer capacitance; (d) solid-contact reference electrode. Reproduced with permission from Ref. \cite{Hu2016}. \textcopyright~2015 Elsevier B.V.}
    \label{fig:solid_ISEs}
\end{figure}

Similar to solid-contact ion-selective electrodes (ISEs), solid-contact reference electrodes play an equally important role in all-solid-state potentiometric systems. These electrodes do not rely on the redox equilibrium of a conventional \ce{Ag}/\ce{AgCl} half-cell to establish a reference potential. Instead, they employ a specially designed reference membrane positioned between the solid-contact layer and the sample solution, forming a stable and reproducible reference potential across the junction between the solid contact and the sample solution, which remains independent of the sample composition. For details regarding its implementation and working mechanism, see Ref. \cite{Lewenstam2013}.

In wearable sweat analysis, potentiometric sensors have been employed for the detection of various ionic analytes, including \ce{Na+} \cite{Gao2016,Emaminejad2017,Bandodkar2014,Anastasova2017,Alizadeh2018,Schazmann2010}, \ce{Cl-} \cite{Emaminejad2017}, \ce{K+} \cite{Gao2016,Alizadeh2018,Sempionatto2017}, \ce{NH4+} \cite{Yu2020}, \ce{Ca^2+} \cite{Nyein2016}, and pH \cite{Anastasova2017,Yu2020,Nyein2016,Bandodkar2013a}, as well as heavy metal ions such as \ce{Cd^2+} \cite{Gupta2014}. Gao \textit{et al}. proposed a fully integrated wearable sensor array for sweat monitoring (see Figure~\ref{fig:pot_designs}a), which represents a milestone in the field of wearable bioanalysis \cite{Gao2016}. In this design, the researchers integrated potentiometric sodium and potassium sensors together with amperometric glucose and lactate sensors onto a single flexible array (see Figure~\ref{fig:pot_designs}b), enabling in situ multiplexed sweat analysis. The sodium and potassium sensors are respectively based on \ce{Na+}- and \ce{K+}-selective electrodes, which share a common solid-state reference electrode. This reference electrode is constructed on an \ce{Ag}/\ce{AgCl} substrate and encapsulated by a poly(vinyl butyral) (PVB) reference membrane. The incorporation of the PVB membrane greatly improves the stability of the solid-state \ce{Ag}/\ce{AgCl} reference electrode and makes it suitable for wearable applications. In $10$--$160~\mathrm{mM}$ \ce{Na+} and $1$--$32~\mathrm{mM}$ \ce{K+} standard solutions, the potentiometric sensors both exhibited near-Nernstian responses, with sensitivities of $64.2~\mathrm{mV/decade}$ for the \ce{Na+} sensor and $61.3~\mathrm{mV/decade}$ for the \ce{K+} sensor (see Figure~\ref{fig:pot_designs}c).

Potentiometry can also be indirectly applied to detect non-ionic molecules such as urea and glucose, which cannot be measured directly by ion-selective electrodes (ISEs). One study reported a biofuel-powered electronic skin (see Figure~\ref{fig:pot_designs}f) that is fully driven by lactate biofuel cells that harvest energy from sweat lactate \cite{Yu2020}. This design incorporates a \ce{NH4+}/urea sensor array and a pH/glucose sensor array, both entirely based on potentiometry. The indicator electrodes of the \ce{NH4+} sensors are \ce{NH4+} ISEs, whereas the urea sensors consist of the same \ce{NH4+} ISEs with an additional urease enzymatic layer (see Figure~\ref{fig:pot_designs}g). Because the decomposition of urea under urease catalysis produces \ce{NH4+}, the ISE can measure the increased \ce{NH4+} concentration in the microenvironment of the urea sensor, thereby indirectly reflecting the urea content in the sample through a potentiometric readout. In parallel, the pH sensors are based on an \ce{H+}-selective ISE that uses an electropolymerized polyaniline film as the ISM. For glucose sensing, the sensing electrode adopts a Nafion/chitosan (CS)--GOx/Nafion sandwich structure constructed on a platinum substrate (see Figure~\ref{fig:pot_designs}h). Although the operating mechanisms were not discussed in detail in the original article, it is reasonable to infer that this electrode operates by detecting the local pH shift induced by the oxidation products of glucose, thereby indirectly reporting the glucose concentration via potentiometry. In the experiments, all four sensors---\ce{NH4+}, urea, pH, and glucose---exhibited a linear relationship between the measured potential and the corresponding analyte concentration (see Figure~\ref{fig:pot_designs}i).


\subsubsection{Amperometric Sensors}
\label{sec:amperometry}
\begin{figure}[h]
    \centering
    \includegraphics[width=0.9\columnwidth]{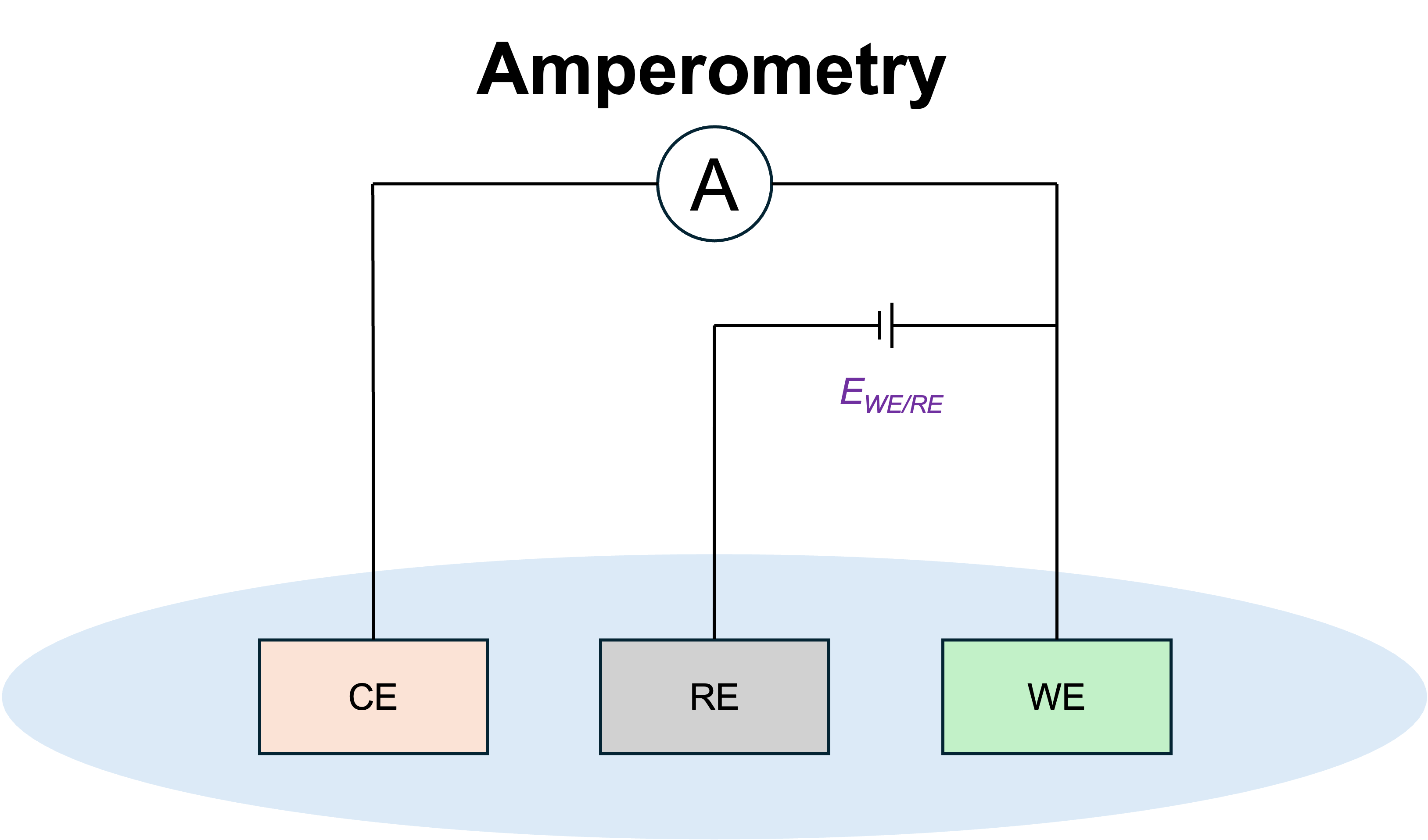}
    \caption{Schematic illustration of an amperometric three-electrode system, where the current between the WE and CE is measured. The current flow through the RE is negligible, and the potential between the WE and RE is held constant by a potentiostat. WE: working electrode; RE: reference electrode; CE: counter electrode.}
    \label{fig:amp_schematic}
\end{figure}

Amperometry \cite{Adeloju2005} is an electrochemical analytical technique that applies a constant potential to an electrode to drive the redox reaction of the target analyte and measures the resulting current to quantify its concentration. The method is suitable for detecting electroactive molecules that can undergo redox reactions under an applied potential, as well as analytes that generate electroactive products through enzymatic catalysis, which can then be detected by amperometry. In contrast to potentiometry, amperometric measurements typically employ a three-electrode system consisting of a reference electrode that provides a stable reference potential, a working electrode at which the redox reaction of interest and signal transduction occur, and an inert counter electrode that collects the electrons transferred during the redox process, thereby forming a closed electrical circuit (see Figure~\ref{fig:amp_schematic}). This is achieved by a potentiostat \cite{Ahmadi2009}, an electronic device that controls the potential difference between the working and reference electrodes. It operates through an actively applied set potential and a negative feedback control loop that continuously adjusts the potential of the counter electrode to compensate for any deviation, thus maintaining a constant working-reference potential while allowing the counter electrode potential to float freely. The implementation of potentiostat will be discussed in detail in Section 4.

When a potential step is applied to the working electrode, the resulting transient current arising from the redox reaction can be described by the Cottrell equation \cite{Adeloju2005}:
\begin{equation}
i(t) = \frac{n F A C_b \sqrt{D}}{\sqrt{\pi t}}
\end{equation}
where $i$ is the transient current, $n$ is the number of electrons transferred in the electrode reaction, $F$ is the Faraday constant, $A$ is the area of the electrode, $C_b$ is the bulk concentration of the target analyte far from the electrode surface (unaffected by the reaction), $D$ is the diffusion coefficient of the analyte, and $t$ is the time elapsed after the potential step. The Cottrell equation describes the scenario in which target molecules are transported from the bulk solution to the electrode surface solely by diffusion during a redox reaction. This model assumes that the reactant on the electrode surface is consumed instantly. As time progresses, a diffusion layer with a concentration gradient of the reactant forms between the electrode surface and the bulk solution. The thickness of this diffusion layer continuously increases, which in turn limits the reaction rate. This phenomenon is reflected in the Cottrell equation as an inverse relationship between the current $i$ and $\sqrt{t}$.

However, in practice, target molecules are not transported to the electrode surface by diffusion alone, as additional transport mechanisms, including convective flow and enzymatic regeneration in enzyme-modified electrodes, continuously replenish electroactive species near the interface. As a result, the diffusion layer does not continue to expand indefinitely but instead reaches a constant effective thickness, under which the current attains its steady state and follows the following empirical relationship:
\begin{equation}
i_{ss} = \frac{n F A C_b D}{\delta}
\label{eq:iss}
\end{equation}
where $\delta$ describes the effective thickness of the steady‐state diffusion layer between the electrode surface and the bulk solution. In biosensor applications, the measured signal typically corresponds to this steady‐state current. Equation~\ref{eq:iss} demonstrates the proportional relationship between the steady‐state current and the analyte concentration in the sample under mass‐transport‐limited conditions, and the specific relationship is experimentally determined by calibration.

To achieve selective detection of target analytes, the working electrode of amperometric biosensors commonly uses enzyme-modified electrode \cite{Updike1967}. The surface of the enzyme-modified electrode is functionalized with specific enzymes that catalyze particular substrates, enabling the selective recognition of the corresponding analytes. Such electrodes have been widely used for the detection of glucose \cite{Gao2014,Ang2015,Cetin2018}, fructose \cite{Ikeda1991,Trivedi2009,Antiochia2013}, lactate \cite{Tu2016,Abrar2016}, alcohols \cite{Choi2007,Kuswandi2014}, aldehydes \cite{Ghica2007,Achmann2008,IndangMarzuki2012}, and polyphenols \cite{CerratoAlvarez2019}. Among these, glucose oxidase (GOx)-based enzyme electrodes were the first to be implemented and are the most extensively investigated \cite{Pinyou2019}. Furthermore, enzyme-modified electrodes can also be used to quantify enzyme activity inhibitors, such as heavy metals \cite{BagalKestwal2008}, pesticides \cite{Arduini2019}, and toxins \cite{Campas2012}, by monitoring the decrease in the electric signal at a fixed substrate concentration. In the operation of enzyme electrodes, the reaction current is limited not only by the mass transport of the reactant but also by the enzymatic kinetics. The Michaelis–Menten equation describes the relationship between the enzymatic reaction rate and the substrate concentration:
\begin{equation}
v = \frac{V_{\max} C_S}{K_M + C_S}
\end{equation}
where $v$ represents the enzymatic reaction rate (in $\mathrm{mol \cdot s^{-1}}$), $V_{\max}$ is the maximum catalytic rate corresponding to the condition in which all active enzyme sites are occupied by the substrate, $C_S$ is the substrate concentration at the enzyme surface, and $K_M$ is the Michaelis constant, defined as the substrate concentration at which the reaction rate reaches half of $V_{\max}$. Incorporating Faraday’s law of electrolysis, the current under enzymatic kinetic control can be expressed as:
\begin{equation}
i = \frac{i_{\text{max}} C_S}{K_M + C_S}
\end{equation}
where $i_{\text{max}} = n F A V_{\text{max}}$. At low substrate concentrations ($C_S \ll K_M$), the current exhibits a linear dependence on the substrate concentration, which can be approximated as:
\begin{equation}
i \approx \frac{i_{\text{max}}}{K_M} C_S
\end{equation}
In practical amperometric biosensors, the operating range is typically confined to this linear regime to ensure a proportional relationship between the substrate concentration and the measured current. The overall current is simultaneously limited by both mass-transport and enzymatic kinetics, which can be described by the following model:
\begin{equation}
\frac{1}{i} = \frac{1}{i_d} + \frac{1}{i_{\text{enz}}}
\end{equation}
where $i$ is the overall measured current, $i_d$ represents the diffusion-limited current accounting for the restrictions of mass-transport, and $i_{\text{enz}}$ denotes the enzyme-kinetic-limited current governed by the catalytic turnover of the enzyme.

\begin{figure}[h]
    \centering
    \includegraphics[width=0.9\columnwidth]{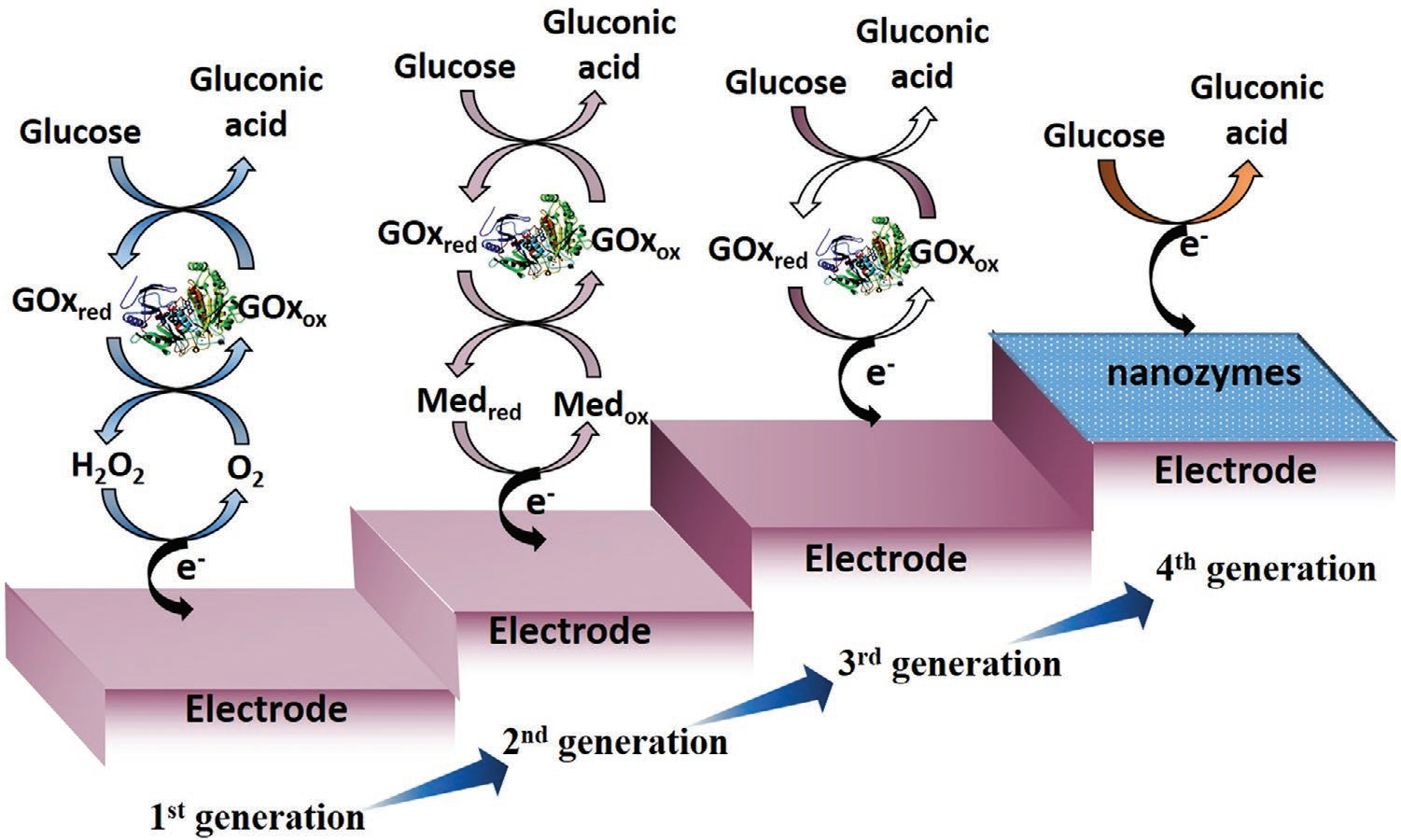}
    \caption{Schematic illustrations of the four generations of enzyme-modified glucose electrodes, shown from left to right as the 1st through 4th generation. Reproduced with permission from Ref. \cite{Adeel2021}. \textcopyright~2021 Wiley‐VCH GmbH.}
    \label{fig:enzyme_electrodes}
\end{figure}

Since the enzyme-modified electrode was first introduced by Leland Clark in 1962, its implementation has been continuously refined and evolved, giving rise to four generations (see Figure~\ref{fig:enzyme_electrodes}) based on progressive improvements in the electron transfer mechanisms \cite{Adeel2021,Nor2022}. The first-generation enzyme electrode is based on a platinum (Pt) electrode modified with glucose oxidase (GOx), operating according to the following reactions:
\begin{align}
\ce{Glucose + O2 &->[GOx] Gluconic acid + H2O2} \\
\ce{H2O2 &->[Pt] 2H+ + O2 + 2e-}
\end{align}
In this system, GOx catalyzes the oxidation of glucose to gluconic acid, utilizing dissolved \ce{O2} in the sample as the natural electron acceptor. The transferred electrons are first stored in the reduced product \ce{H2O2}, which is subsequently oxidized on the Pt electrode to generate the measurable current. Thus, the system does not directly detect glucose oxidation itself, but instead measures the oxidation current of \ce{H2O2}, indirectly reflecting the glucose concentration in the sample. Because the oxidation current of \ce{H2O2} is proportional to its generation rate, which in turn directly reflects the glucose oxidation rate, the measured current can be quantitatively correlated with the glucose concentration in the sample. However, the major drawback of this design lies in its strong dependence on the dissolved oxygen in the sample. The solubility of oxygen in biological fluids is inherently low, which can lead to the electrode reaction being limited by oxygen availability rather than by the concentration of the target analyte. Furthermore, fluctuations in the local oxygen concentration can cause significant current drift, thus reducing the accuracy and reliability of the glucose measurement \cite{Adeel2020,Chen2013a,Heller2008}.

\begin{figure*}[t]
    \centering
    \includegraphics[width=0.9\textwidth]{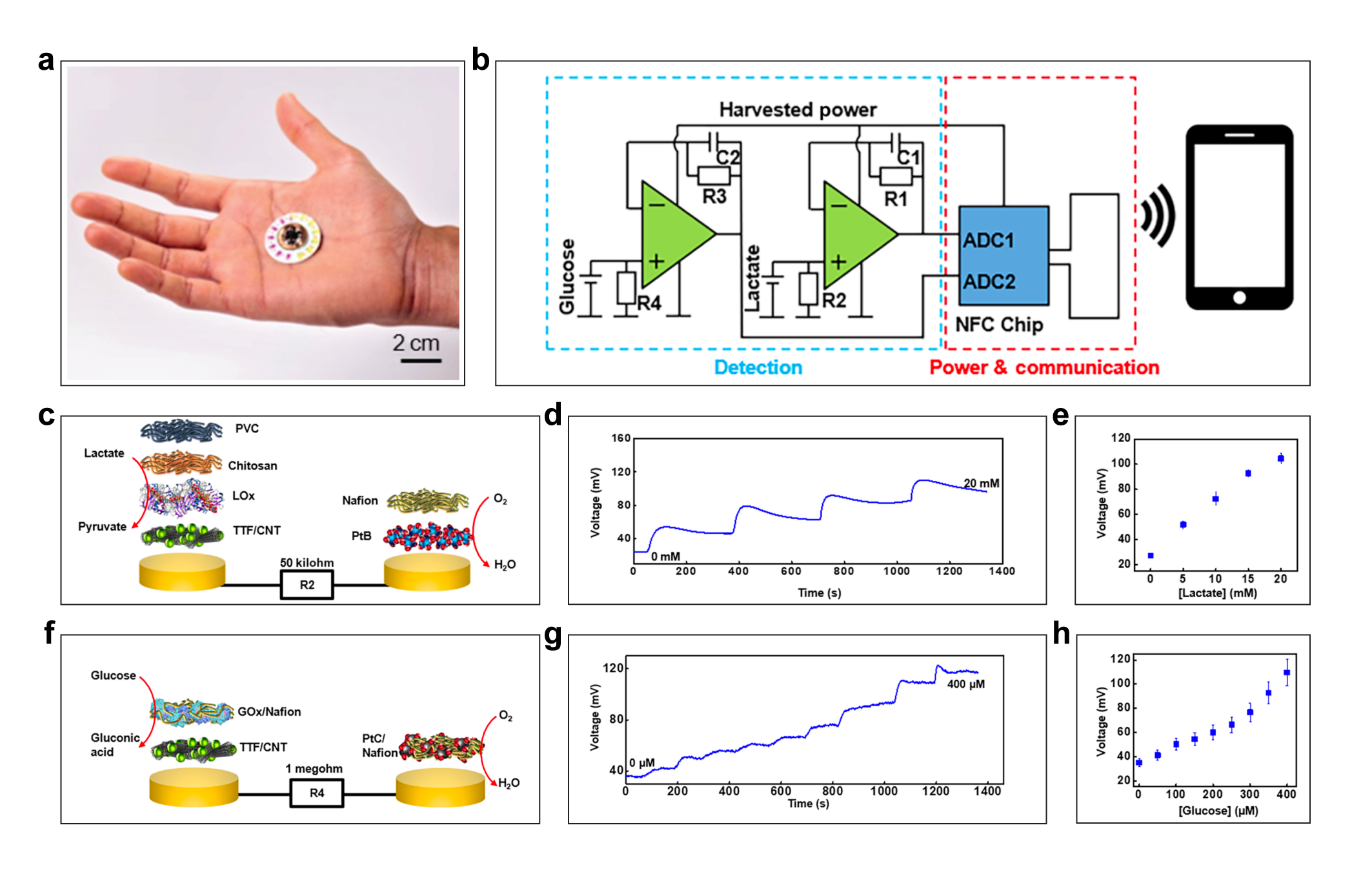}
    \caption{A battery-free, skin-mounted microfluidic/electronic platform for simultaneous electrochemical and colorimetric sweat analysis. (a) Photograph of the device held in the hand. (b) Schematic diagram of the system for signal readout and wireless transmission. (c) Schematic illustration of the lactate sensor. (d,e) Real-time lactate sensing response and the corresponding calibration curve. (f) Schematic illustration of the glucose sensor. (g,h) Real-time glucose sensing response and the corresponding calibration curve. Reproduced from Ref.~\cite{Bandodkar2019b}.}
    \label{fig:amp_designs}
\end{figure*}

To overcome the dependence of the system on dissolved oxygen, second-generation enzyme electrodes introduced artificial reversible redox mediators such as ferrocene derivatives, ferricyanide/ferrocyanide \cite{Chen2013a,Heller2008,Lee2018}, and quinone compounds \cite{Chen2013a,Heller2008}, which replace \ce{O2} as the electron acceptors. These mediators diffuse to the enzyme active site, accept electrons generated during substrate oxidation, and are subsequently reoxidized at the electrode surface, thereby shuttling electrons between the enzyme and the electrode. Compared with the first generation, this design effectively eliminates the dependence on \ce{O2} and significantly lowers the operating potential, as the redox potentials of the artificial mediators are much lower than that of \ce{H2O2}. However, this generation still exhibits several limitations, including the poor stability and potential leakage of the mediators, as well as the continued lack of direct electron transfer between the enzyme and the electrode.

The third-generation enzyme electrodes represent a major conceptual breakthrough in the evolution of enzyme-based electrochemical sensors, as they eliminate the need for electron mediators and enable direct electron transfer (DET) between the enzyme and the electrode. The major challenge in achieving DET lies in minimizing the distance between the redox center of the enzyme and the electrode surface. During the catalytic oxidation of glucose, the cofactor \ce{FAD} in glucose oxidase (GOx) is reduced to \ce{FADH2}. However, the \ce{FAD}/\ce{FADH2} redox couple is deeply embedded within the protein structure, and the tunneling efficiency of electrons decays exponentially with increasing distance, preventing efficient electron transfer to the electrode \cite{Chen2013a}. Thanks to the utilization of nanomaterials, various strategies have been developed to bridge this spatial gap. For example, nanostructured electrode surfaces can be employed to promote intimate contact between the enzyme and the electrode, thereby physically reducing the distance between the \ce{FAD} center and the electrode interface \cite{Adeel2020,Lee2018}, among several other approaches.

Recently, electrode platforms leveraging nanomaterials with enzyme-mimicking catalytic activity, commonly known as nanozymes, have been developed. These electrodes, known as fourth-generation enzyme electrodes, achieve catalytic oxidation of the target analyte through intrinsic redox-active sites within the nanomaterials, thereby eliminating the dependence on natural enzymes. Because these electrodes are more robust and exhibit better long-term stability than natural enzyme-based electrodes, they are particularly well suited for integration into wearable and miniaturized sensing systems.

In wearable sweat monitoring platforms, amperometric sensors are widely employed for the detection of glucose \cite{Gao2016,Emaminejad2017,Bandodkar2019b,Zhao2019}, lactate \cite{Gao2016,Anastasova2017,Sempionatto2017,Bandodkar2019b,Jia2013}, alcohol \cite{Kim2016,Hauke2018}, ascorbic acid \cite{Zhao2020,Sempionatto2020}, levodopa \cite{Tai2019}, and nicotine \cite{Tai2020}. The amperometric system of the design proposed by Gao \textit{et al}. adopts a two-electrode configuration, in which the glucose and lactate working electrodes share the same \ce{Ag}/\ce{AgCl} electrode that simultaneously functions as the reference and counter electrode. Although the reference potential in a two-electrode amperometric system is generally less stable than in a three-electrode configuration, the extremely low current levels required in this system allow such a simplified architecture without compromising measurement accuracy. The glucose and lactate working electrodes are variants of first-generation enzyme electrodes based on glucose oxidase (GOx) and lactate oxidase (LOx), respectively. Both electrodes employ \ce{Cr}/\ce{Au} films as the conductive substrate and incorporate an intermediate Prussian blue (PB) layer between the enzyme layer and the metal base. PB acts as an additional electrocatalytic mediator that facilitates electron transfer between enzymatically generated \ce{H2O2} and the electrode surface, which lowers the operational potential for \ce{H2O2} reduction to $\sim 0~\text{V}$. Furthermore, the thickness of the PB layer can be tuned to tailor the sensor’s response characteristics: a thinner PB film is used for the glucose sensor to enhance sensitivity in the low-concentration regime, whereas a thicker PB layer is adopted for the lactate sensor to maintain linearity across the relatively high lactate concentrations typically found in sweat. In the $0$--$200~\mu\mathrm{M}$ glucose and $0$--$30~\mathrm{mM}$ lactate solutions, the chronoamperometric current responses of both sensors exhibited linear current--concentration relationships after calibration (see Figure~\ref{fig:pot_designs}c). The glucose sensor showed a sensitivity of $2.35~\mathrm{nA \cdot \mu M^{-1}}$, while the lactate sensor exhibited a sensitivity of $220~\mathrm{nA \cdot mM^{-1}}$.

In addition, Bandodkar \textit{et al}. proposed a wearable sweat-sensing platform powered wirelessly via near-field communication (NFC) technology (see Figure~\ref{fig:amp_designs}a), which can detect lactate and glucose in sweat based on the principle of biofuel cells (BFCs) \cite{Bandodkar2019b}. BFCs and traditional amperometric methods both rely on redox reactions of the target analytes to generate electrical currents, which exhibit a positive correlation with the analyte concentration within their linear operating range. The key difference is that BFCs generate a driving potential through spontaneous reactions between the anode and the cathode, eliminating the need for an externally applied constant bias voltage and thus the potentiostat, thereby enabling fully self-powered electrochemical sensing. The anode of the lactate BFC uses a carbon nanotube (CNT) paper substrate mixed with tetrathiafulvalene (TTF) as the electron mediator, on top of which a LOx layer is immobilized (see Figure~\ref{fig:amp_designs}c). Because the concentration of lactate in sweat is approximately $14~\mathrm{mM}$ \cite{Harvey2010}, much higher than the Michaelis constant of LOx ($\sim 1~\mathrm{mM}$), a chitosan layer and a polyvinyl chloride (PVC) layer are coated onto the anode to prevent saturation of the enzyme layer and ensure that the response remains within the linear regime. These layers form a tight diffusion barrier that reduces the rate at which lactate in the sample reaches the enzyme layer, while simultaneously minimizing the leakage of mediators. To read out the generated current signal, a $50~\text{k}\Omega$ load resistor is connected between the output terminal of the lactate BFC and the reference ground, which linearly converts the current signal into a corresponding voltage signal. The implementation of the glucose sensor follows a similar configuration (see Figure~\ref{fig:amp_designs}f); however, due to the much lower glucose concentration in sweat, the enzyme layer is formed by directly dispersing GOx within Nafion, which ensures a fast interaction between glucose and the enzyme. Because the resulting currents are smaller, a load resistor of $1~\text{M}\Omega$ is selected for the glucose BFC to generate a sufficiently high voltage signal.


\subsubsection{Voltammetric Sensors}
Voltammetry is an electrochemical analytical technique that applies a specific time-dependent potential waveform to an electrode, and records the relationship between the redox current as a function of the electrode potential at each moment to investigate the electrochemical behavior of the system. Similar to amperometry, a voltammetric system also employs a three-electrode configuration, consisting of one electrode to which the potential waveform is applied, a counter electrode that completes the current loop, and a reference electrode that provides a stable reference potential. The key difference between the two lies in the applied potential, which remains constant in amperometry but varies with time in voltammetry. In electrochemical sensing, voltammetric techniques can be broadly categorized according to their purposes into mechanistic (or diagnostic) voltammetry and analytical (or sensing) voltammetry. The former is used to investigate and understand the fundamental properties of electrochemical reaction systems, whereas the latter, similar to potentiometry and amperometry, is utilized for the quantitative detection of analyte concentrations. To meet different analytical and mechanistic demands, various voltammetric modes with distinct potential waveforms and scanning strategies have been developed. Among them, the most relevant techniques for wearable biosensing applications are linear sweep voltammetry (LSV), cyclic voltammetry (CV), differential pulse voltammetry (DPV), and square wave voltammetry (SWV).

The idea of linear sweep voltammetry (LSV) is to scan the potential within a fixed range from a lower limit to an upper limit at a constant rate, defined as $v = dE/dt$. When the potential is below the onset potential of the target reaction, the current in the system mainly originates from unrelated background processes and remains nearly constant or changes only slightly with potential. As the potential reaches the onset potential, the current begins to increase rapidly with the potential and gradually approaches the limiting current that can be achieved under mass-transport limitation. The limiting current is also described by Equation~\ref{eq:iss}. In the application of wearable sweat sensors, LSV is typically used to identify the appropriate operating potential for subsequent amperometric detection by examining the onset potential and the limiting current region in the current–potential curve \cite{Jia2013}. In other applications, LSV can also be employed for quantitative analysis. In such cases, LSV measurements are carried out at a series of known substrate concentrations, and the resulting peak currents are used to construct a calibration curve, since steady-state limiting currents are typically difficult to achieve in LSV. The analyte concentration can then be quantitatively determined on the basis of this calibration curve \cite{Chelly2021,Hussain2017}.

Cyclic voltammetry (CV) \cite{Kissinger1983} is an extension of linear sweep voltammetry (LSV). In CV, the potential is linearly swept within a fixed potential window from a lower to an upper limit and then reversed back to the initial potential. During the forward and reverse scans, the system undergoes oxidation and reduction processes, respectively, typically exhibiting a pair of opposite anodic and cathodic peaks on the current–potential curve. Abundant electrochemical information can be obtained by analyzing the peak currents and peak potentials, including the reaction reversibility, kinetics, and the electrochemical activity and stability of electrode modification layers. In wearable sweat sensors, CV is commonly used to evaluate the electrochemical performance of the sensor and to analyze the redox characteristics of the system in order to determine the appropriate operating potential \cite{Zhao2019,Kim2016,Hauke2018,Zhao2020,Tai2019,Tai2020}. Beyond characterization, CV can also serve as a controllable electrochemical deposition method for sensor fabrication, where the alternating oxidation and reduction cycles at the electrode surface progressively drive material deposition or film growth \cite{Zhao2019,Tai2019}.

\begin{figure*}[t]
    \centering
    \includegraphics[width=0.9\textwidth]{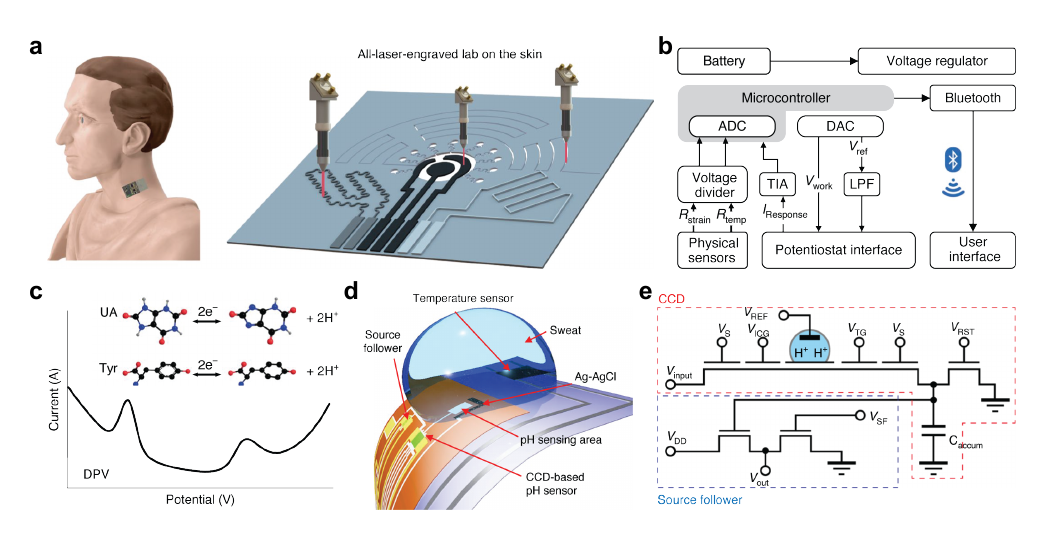}
    \caption{A laser-engraved sensor for sweat UA and Tyr detection (a–c) and a sensor using an FET-based CCD for sweat pH detection (d, e). (a) Schematic illustration of the device on the skin and the laser-engraved components, including the microfluidic module and the physical and chemical sensors. (b) Block diagram of the system showing the power supply, signal readout, and wireless communication with the user interface. (c) Schematic illustration of the DPV response for detecting UA and Tyr. The analyte level is reflected in the height of the oxidation peak. Reproduced from Ref.~\cite{Yang2019}. (d) Schematic of the CCD-based sensor. (e) Schematic illustration of the CCD circuit. Reproduced from Ref.~\cite{Nakata2018}. UA: uric acid; Tyr: tyrosine; CCD: charge-coupled device.}
    \label{fig:vol_fet_designs}
\end{figure*}

Differential Pulse Voltammetry (DPV) is a pulse-based potential scanning technique. It superimposes a series of potential pulses onto a linear potential ramp and measures the difference between the currents recorded before and after each pulse to construct the current–potential curve. This differential measurement effectively reduces the contribution of the background capacitive current, thereby increasing the signal-to-noise ratio and enabling a lower detection limit. Meanwhile, because each pulse lasts for only a short duration, the resulting transient current is less affected by diffusion limitation, leading to a stronger response to analyte concentration and thus a higher sensitivity. These advantages make DPV particularly suitable for trace-level analyte detection. In DPV, sharp peaks appear at the redox potentials of analytes. When multiple electroactive species are present in a system, their individual redox peaks can usually be clearly distinguished, which is particularly useful for the simultaneous detection of multiple analytes and for minimizing interference from non-target species. The peak current of each analyte is linearly proportional to its concentration, providing the basis for quantitative analysis.

A representative study developed a laser-engraved wearable sweat sensing platform that used differential pulse voltammetry (DPV) to detect ultralow concentrations of uric acid (UA) and tyrosine (Tyr) in human sweat. Two distinct oxidation peaks appeared in the current–potential curves, corresponding to the oxidation of UA at $\sim 0.39~\text{V}$ and Tyr at $\sim 0.64~\text{V}$. The sensor exhibited high sensitivities of $3.50~\mu\text{A}\,\mu\text{M}^{-1}\,\text{cm}^{-2}$ for UA and $0.61~\mu\text{A}\,\mu\text{M}^{-1}\,\text{cm}^{-2}$ for Tyr, with detection limits of $0.74~\mu\text{M}$ and $3.6~\mu\text{M}$, respectively. In addition, the device demonstrated excellent selectivity against common interfering species such as glucose, urea, dopamine, and ascorbic acid \cite{Yang2019}. In addition, another study has reported the application of DPV in wearable sweat sensors for the direct detection of tryptophan (Trp), as well as for the indirect quantification of other non-electroactive amino acids through the assistance of redox-active reporters (RARs) \cite{Wang2022b}. Trace-level drugs present in sweat, such as caffeine, can also be quantitatively monitored in real time using DPV \cite{Tai2018}. Beyond wearable sensors, DPV has been demonstrated to be capable of detecting trace-level neuropeptide Y (NPY) in buffered samples \cite{Sanghavi2014}, highlighting its potential applicability to sweat-based NPY monitoring.

Square-wave voltammetry (SWV) \cite{Chen2013b} is another potential scanning technique based on superimposing a series of square-wave pulses onto a baseline. Although the underlying idea is to some extent similar to DPV, SWV employs a different scanning waveform and measurement scheme, which makes it preferable in certain scenarios. In SWV, the baseline is a staircase potential, and a pair of positive and negative square-wave pulses is superimposed onto each step. The currents are sampled near the end of the positive and negative pulses, respectively, at which point the capacitive charging currents have already decayed during the pulse plateau, and their difference is recorded. This differential current is then plotted as a function of the applied potential to obtain the SWV curve. Compared with DPV, the most significant advantage of SWV is its markedly higher scanning speed. This is because the forward and reverse pulses in SWV are sampled at the same potential stair, where the capacitive background currents are mirror-symmetric and therefore more effectively canceled upon differentiation, eliminating the need to wait at each step for the capacitive transient to decay and for the diffusion layer to re-establish a reproducible state. Moreover, SWV generally offers higher sensitivity than DPV, and its $I$--$V$ graph often exhibits sharper peak shapes. However, despite these advantages, SWV is less widely employed for quantitative analysis. This is primarily because the peak current in SWV is highly sensitive to experimental conditions, particularly waveform parameters and interfacial kinetics, which makes it difficult to maintain a stable current--concentration relationship across measurements. As a result, its reproducibility is typically inferior to that of DPV. In practice, SWV is typically applied for rapid on-site screening, qualitative analysis, and semi-quantitative analysis \cite{Mishra2017,Jong2016,Bandodkar2013b}. A study has demonstrated the capability of SWV for wearable pharmacokinetic monitoring \cite{Lin2022}.

\subsubsection{Field-Effect Transistor (FET)--Based Electrochemical Sensors}
Field-Effect Transistors (FETs) play essential roles in modern analog and digital circuits. In analog applications, their most important property is the ability to adjust the carrier density in the channel by modifying the gate–source voltage, which in turn changes the channel conductance. When the gate--source voltage ($V_{GS}$) is lower than the threshold voltage ($V_{th}$), the device remains in the cut-off state. Once $V_{GS} > V_{th}$, the operating region is determined by the relationship between the overdrive voltage ($V_{ov} = V_{GS} - V_{th}$) and the drain--source voltage ($V_{DS}$). The transistor operates in the triode region when $V_{ov} > V_{DS}$, and in the saturation region when $V_{ov} \le V_{DS}$. The corresponding drain current expressions (for an n-type device and neglecting the Early effect) are:
\begin{align}
I_{D,\text{tri}} &= \mu_n C_{ox}\frac{W}{L}
\left[(V_{GS}-V_{th})V_{DS} - \frac{V_{DS}^2}{2}\right] \\
I_{D,\text{sat}} &=
\frac{1}{2}\mu_n C_{ox}\frac{W}{L}(V_{GS}-V_{th})^2
\end{align}
Here, $I_{D,\text{tri}}$ and $I_{D,\text{sat}}$ represent the drain currents in the triode and saturation regions, respectively; $\mu_n$ denotes the electron mobility in the channel; and $C_{ox}$, $W$, and $L$ are structural parameters of the device, corresponding to the gate oxide capacitance per unit area, the channel width, and the channel length. In analog circuit design, MOSFETs are typically operated in the saturation region. When $V_{GS}$ is biased at a fixed operating point and a small perturbation $v_{gs}$ is superimposed on this bias, the resulting incremental drain current can be modeled as a current source controlled by $v_{gs}$, which is called the small-signal model. The small-signal current can be expressed as:
\begin{align}
i_{d} &= g_m v_{gs} \\
g_m &= \frac{\partial i_{D}}{\partial v_{GS}} \; @ V_{GS}
\end{align}
where $g_m$ is called transconductance, which is defined as the partial derivative of the static drain current with respect to the static gate--source voltage at the bias point.

The ability of FETs to amplify small perturbations applied at the gate terminal has inspired their extensive use in biosensors \cite{Li2020}. The core idea of FET--based electrochemical biosensors is to couple the concentration of the target analyte in sample to the gate potential of the transistor. Variations in the effective gate--source voltage can then be transduced into measurable electrical outputs, typically changes in the drain current. In practice, by employing analyte-sensitive materials, the presence of the target analyte can introduce a shift superimposed on the constant gate bias, essentially a boundary potential directly related to its concentration. This perturbation modulates the carrier concentration (and hence the conductivity) of the transistor channel, and the resulting change manifests as a variation in the drain current \cite{Liu2018,GarciaCordero2018,Liao2014,Minami2015}. Alternatively, charge-based transduction can be employed, in which the analyte concentration modulates the potential of the sensing gate, which in turn defines the depth of the potential well formed beneath it. When an input-gate injection pulse is applied, electrons are captured in this potential well, and the amount of charge captured is dictated by its depth. The resulting charge packets are subsequently transferred through a series of transfer gates and ultimately accumulated on an output capacitor, where a voltage proportional to the total stored charge is generated. By reading this capacitor voltage, the analyte concentration can be quantitatively determined \cite{Nakata2018}. It is also possible, without changing the gate--source voltage, to directly dope the channel with electrons released from the analyte’s oxidation, thereby altering its conductance and reflecting the analyte concentration through the corresponding change in the drain current \cite{Pappa2018}. In wearable sweat-sensing applications, FET-based sensors have been used to detect \ce{Na+} and \ce{K+} \cite{GarciaCordero2018}, pH \cite{GarciaCordero2018,Nakata2018}, glucose \cite{Liu2018}, and lactate \cite{Minami2015,Pappa2018}.

\subsection{Non-Electrochemical Sensors}
In addition to electrochemical sensors, a variety of sensor types based on non-electrochemical mechanisms have also been widely employed in wearable sweat sensing, providing diversified options for analyte detection. Among these, optical sensors are one of the most common categories. The major types of optical sensors include absorption-based sensors, fluorescence and luminescence sensors, scattering-based sensors, and interferometric or refractive-index sensors \cite{Min2023}. Among them, colorimetry, which is an absorption-based method, is by far the most widely utilized technique in wearable sweat sensors. Wearable colorimetric sensors are often integrated with microfluidic platforms. A representative design introduced a soft, flexible, and skin-conformal microfluidic patch capable of performing multiplexed sweat analysis through colorimetry \cite{Koh2016}. Sweat is routed through a serpentine microchannel and subsequently into four spatially isolated reservoirs, each preloaded with cellulose-based substrates containing immobilized reagents for the colorimetric detection of pH, \ce{Cl-}, lactate, and glucose. The device also incorporates a battery-free NFC module that wirelessly triggers smartphone-based image capture and analysis when brought into proximity, enabling quantitative readout of the colorimetric responses. Another study reported a wearable colorimetric sweat glucose sensor based on a microfluidic chip \cite{Xiao2019}. Sweat is passively routed through microfluidic channels into reaction chambers preloaded with glucose oxidase--peroxidase--o-dianisidine reagents, where the colorimetric reaction takes place. The resulting color change can be qualitatively evaluated by the naked eye or quantitatively analyzed through smartphone imaging coupled with color‐difference analysis. In addition, electrochemistry-based wearable sweat sensors can be augmented with colorimetric elements, allowing optical quantification of parameters such as pH and chloride in parallel with electrochemical detection \cite{Bandodkar2019b}.

Another class of wearable sensors is based on the piezoelectric effect. Scarpa \textit{et al}. developed a wearable-capable pH sensor that exploits the inverse piezoelectric effect, in which an applied electric field induces mechanical vibration in the resonator \cite{Scarpa2020}. The device consists of four identical aluminum nitride (AlN) microbalances acting as piezoelectric resonators, each patterned with a pH-responsive PEG-DA/CEA hydrogel. The AlN layer is sandwiched between two molybdenum electrodes, and an AC voltage applied across these electrodes generates an oscillating electric field that drives the membranes into resonance. The carboxyl groups (\ce{-COOH}) in the hydrogel undergo reversible protonation and deprotonation depending on pH; deprotonation increases the fraction of \ce{-COO^-} groups, enhancing electrostatic repulsion and causing the hydrogel to swell and absorb water, thereby increasing its mass, and vice versa. The resulting mass changes induce downshifts or upshifts in the resonant frequency, enabling quantitative determination of pH by monitoring the resonance frequency shift. Another design utilizes the direct piezoelectric effect for sweat analysis, in which mechanical deformation induces an electrical signal \cite{Han2017b}. The core sensing element consists of enzyme-modified \ce{ZnO} nanowire piezo-biosensing units bridged between interdigital electrodes. When external mechanical deformation is applied to the flexible substrate, the strain is transferred to the ZnO nanowires, causing them to bend along their c-axis and generate transient piezoelectric polarization, which appears as measurable piezoelectric impulses. To enable analyte detection, each nanowire is functionalized with a specific enzyme: lactate oxidase, glucose oxidase, uricase, or urease. The corresponding enzymatic reactions generate \ce{H+}, \ce{e^-}, or \ce{NH4+}, which adsorb onto the nanowire surface, increasing the surface carrier density and thereby enhancing piezoelectric field screening. As a result, under the same mechanical deformation, the amplitude of the piezoelectric impulse decreases with increasing analyte concentration. By monitoring these amplitude variations, the device enables quantitative detection of lactate, glucose, uric acid, and urea in sweat.

The last category of sensors discussed here are those based on biorecognition. In nature, a wide range of biomolecular interactions exhibit intrinsic specificity, including enzyme--substrate pairs, antigen--antibody binding, nucleic-acid hybridization between complementary strands, and receptor--ligand interactions, among others. Because many disease-related biomarkers in sweat are present at extremely low concentrations, developing sensors with strong molecular specificity that supports ultrahigh sensitivity is essential \cite{Tu2019}. Sensors based on biorecognition typically consist of a bioreceptor layer, which provides selective recognition of the target analyte, and a signal-transduction layer, which converts the biorecognition event into a measurable output \cite{Min2023}. To date, a variety of recognition elements, including antibodies, nucleic acids, receptor proteins, and several classes of biomimetic materials, have been incorporated into wearable sweat sensors \cite{Morales2018}. As an example, immunosensors based on antigen--antibody recognition have been adapted for wearable sweat analysis and have been employed to detect a variety of trace biomarkers, including cortisol \cite{Upasham2020b,Munje2017,Upasham2022,Kinnamon2017}, dehydroepiandrosterone (DHEA) \cite{Upasham2020b}, cytokines \cite{Jagannath2020,Munje2017,Upasham2022,Jagannath2021,Kothari2022}, C-reactive protein (CRP) \cite{Jagannath2020}, neuropeptide Y (NPY) \cite{Churcher2020}, and ethyl glucuronide (ETG, a direct metabolite of ethanol) \cite{Selvam2016}. 



\bibliographystyle{unsrt}
\bibliography{references}

\end{document}